\documentclass[aps,prc,preprint,superscriptaddress,groupedaddress,
showpacs,floatfix,amsmath,amssymb,]{revtex4}
\usepackage{graphicx}

\begin{document}

\title{Superscaling analysis of inclusive electron scattering
and its extension to charge-changing neutrino-nucleus cross
sections beyond the relativistic Fermi gas approach}

\author{A.~N.~Antonov}
\affiliation{Institute for Nuclear Research and Nuclear Energy,
Bulgarian Academy of Sciences, Sofia 1784, Bulgaria}

\author{M.~V.~Ivanov}
\affiliation{Institute for Nuclear Research and Nuclear Energy,
Bulgarian Academy of Sciences, Sofia 1784, Bulgaria}

\author{M.~K.~Gaidarov}
\affiliation{Institute for Nuclear Research and Nuclear Energy,
Bulgarian Academy of Sciences, Sofia 1784, Bulgaria}

\author{E.~Moya de Guerra}
\affiliation{Instituto de Estructura de la Materia, CSIC, Serrano
123, 28006 Madrid, Spain}

\affiliation{Departamento de Fisica Atomica, Molecular y Nuclear,\\
Facultad de Ciencias Fisicas, Universidad Complutense de Madrid,
Madrid E-28040, Spain}

\author{J.~A.~Caballero}
\affiliation{Departamento de Fisica Atomica, Molecular y Nuclear,\\
Universidad de Sevilla, Apdo. 1065, 41080 Sevilla, Spain}

\author{M.~B.~Barbaro}
\affiliation{Dipartimento di Fisica Teorica, Universit\`{a} di
Torino and INFN,\\
Sezione di Torino, Via P. Giuria 1, 10125 Torino, Italy}

\author{J.~M.~Udias}
\affiliation{Departamento de Fisica Atomica, Molecular y Nuclear,\\
Facultad de Ciencias Fisicas, Universidad Complutense de Madrid,
Madrid E-28040, Spain}

\author{P.~Sarriguren}
\affiliation{Instituto de Estructura de la Materia, CSIC, Serrano
123, 28006 Madrid, Spain}

\begin{abstract}
Superscaling analyses of inclusive electron scattering from nuclei
are extended from the quasielastic processes to the delta
excitation region. The calculations of $(e,e^\prime)$ cross
sections for the target nucleus $^{12}$C at various incident
electron energies are performed using scaling functions
$f(\psi^{\prime})$ obtained in approaches going beyond the
mean-field approximation, such as the coherent density fluctuation
model (CDFM) and the one based on the light-front dynamics (LFD)
method. The results are compared with those obtained using the
relativistic Fermi gas (RFG) model and the extended RFG model
(ERFG). Our method utilizes in an equivalent way both basic
nuclear quantities, density and momentum distributions, showing
their role for the scaling and superscaling phenomena. The
approach is extended to consider scaling function for medium and
heavy nuclei with $Z\neq N$ for which the proton and neutron
densities are not similar. The asymmetry of the CDFM quasielastic
scaling function is introduced, simulating in a phenomenological
way the effects which violate the symmetry for $\psi^{\prime}\geq
0$ including the role of the final-state interaction (FSI). The
superscaling properties of the electron scattering are used to
predict charge-changing neutrino-nucleus cross sections at
energies from 1 to 2 GeV. A comparison with the results of the
ERFG model is made. The analyses make it possible to gain
information about the nucleon correlation effects on both local
density and nucleon momentum distributions.
\end{abstract}
\maketitle

\section{Introduction}

Over the past four decades electron scattering has provided a
wealth of information on nuclear structure and dynamics. Form
factors and charge distributions have been extracted from elastic
scattering data, while inelastic measurements have allowed for a
systematic study of the dynamic response over a broad range of
momentum and energy transfer. The nuclear $y$-scaling analysis of
inclusive electron scattering from a large variety of nuclei (e.g.
\cite{West75,Sick80,Day90,Cio87,Cio90,Cio96,AMD+88,CW97,CW99,FCW00})
showed the existence of high-momentum components in the nucleon
momentum distributions $n(k)$ at momenta $k>2$ fm$^{-1}$ due to
the presence of nucleon-nucleon (NN) correlations. It was shown
(see, e.g. \cite{AGK+04,AGK+05,AIG+06,DS99l,DS99}) that this
specific feature of $n(k)$, which is similar for all nuclei, is a
physical reason for the scaling and superscaling phenomena in
nuclei.

The concepts of scaling \cite{West75,Sick80,Day90,Cio87,Cio90,
Cio96,CW97,CW99,FCW00} and superscaling
\cite{AMD+88,AGK+05,AIG+06,DS99l,AGK+04,DS99,BCD+98} have been
explored in \cite{DS99,MDS02} for extensive analyses of the
$(e,e^{\prime})$ world data (see also \cite{Benhar2006}). Scaling
of the first kind (no dependence on the momentum transfer) is
reasonably good as expected, at excitation energies below the
quasielastic (QE) peak, whereas scaling of second kind (no
dependence on the mass number) is excellent in the same region.
When both types of scaling behavior occur one says that
superscaling takes place. At energies above the QE peak both
scaling of the first and, to a lesser extent, of the second kind
are shown to be violated because of important contributions
introduced by effects beyond the impulse approximation, namely,
inelastic scattering \cite{BCD+04,AlvRuso03} together with
correlation contributions and meson exchange currents
\cite{Amaro2002,Pace03}.

The superscaling analyses of inclusive electron scattering from
nuclei for relatively high energies (several hundred MeV to a few
GeV) have recently been extended to include not only quasielastic
processes, but also the region where $\Delta$-excitation dominates
\cite{Amaro2005}. A good representation of the electromagnetic
response in both quasielastic and $\Delta$ regions has been
obtained using the scaling ideas, importantly, with an asymmetric
QE scaling function $f^{QE}(\psi^{\prime})$ ($\psi^{\prime}$ is
the scaling variable in the QE region) and a scaling function
$f^{\Delta}(\psi^{\prime}_{\Delta})$ in the region up to
inelasticities where the $\Delta$ contribution reaches its
maximum. Both functions were deduced from phenomenological fits to
electron scattering data. Particularly, for the scaling function
in the quasielastic region it has been shown in
Ref.~\cite{Amaro2005} that, in contrast to the relativistic Fermi
gas model scaling function, which is symmetric, limited strictly
to the region $-1 \leq \psi^{\prime} \leq +1$, and with a maximum
value 3/4, the empirically determined $f^{QE}(\psi^{\prime})$ has
a somewhat asymmetric shape with a tail that extends towards
positive values of $\psi^{\prime}$ and its maximum is only about
0.6. Of course, the specific features of the scaling function
should be accounted for by reliable microscopic calculations that
take FSI into account. In particular, the asymmetric shape of
$f^{QE}$ tested in Refs.~\cite{Caballero2005,Caballero2006} by
using a relativistic mean field (RMF) for the final states shows a
very good agreement with the behavior presented by the
experimental scaling function.

The superscaling analyses and the present knowledge of inclusive
electron scattering allowed one to start studies of neutrino
scattering off nuclei on the same basis. The reactions of incident
neutrino beams interacting with a complex nucleus have offered
unique opportunities for exploring fundamental questions in
different domains in physics. Recently, positive signals of
neutrino oscillations confirmed the hypothesis of non-zero
neutrino masses and triggered much interest on this issue
\cite{Fukuda}. To better analyze the next generation of
high-precision neutrino oscillation experiments and to reduce
their systematic uncertainty both neutral- (e.g.
\cite{Meucci20041,Amaro2006,Martinez2006,Barbaro2006,Nieves2005})
and charged-current (e.g.
\cite{Amaro2005,Caballero2005,Martinez2006,Barbaro2006,Barbaro2005,
Maieron2003,Meucci20042,Benhar20041,Benhar2004,Benhar2005})
neutrino-nucleus scattering have stimulated detailed
investigations.

The neutrino-nucleus interactions have been studied within several
approaches investigating a variety of effects. Using the
superscaling analysis of few-GeV inclusive electron-scattering
data, a method was proposed in Ref.~\cite{Amaro2005} to predict
the inclusive $\nu A$ and $\overline\nu A$ cross sections for the
case of $^{12}$C in the nuclear resonance region, thereby
effectively including delta isobar degrees of freedom. It was
shown in Refs.~\cite{Meucci20041,Meucci20042} that the important
final-state interaction effects arising from the use of
relativistic optical potentials within a relativistic Green's
function approach lower the cross section by at least a 14\%
factor at incoming neutrino energies of 1 GeV. A similar result
has been obtained in Refs.~\cite{Co2006,Botrugno2005} where the
use of Random Phase Approximation (RPA) to predict the
neutrino-nucleus cross section was discussed. Apart from
relativistic dynamics and FSI, other effects may influence the
neutrino-nucleus reactions. The role of Pauli blocking and FSI in
charged-current neutrino induced reactions is analyzed in
Refs.~\cite{Benhar20041,Benhar2004,Benhar2005}.

In this article we follow our method presented in
Refs.~\cite{AGK+04,AGK+05,AIG+06} to calculate the scaling
function in finite nuclei firstly within the coherent density
fluctuation model (e.g., Refs.~\cite{AHP88,AHP93,ANP79+,A+89+}).
This approach, which is a natural extension of the RFG model, has
shown how both basic quantities, density and momentum
distributions, are responsible for the scaling and superscaling
phenomena in various nuclei. Although the scaling function
obtained in \cite{AGK+04} is symmetrical around $\psi^{\prime}=0$,
the results agree with the available experimental data at
different transferred momenta and energies below the quasielastic
peak position, showing superscaling for $\psi^{\prime}<0$
including $\psi^{\prime}<-1$, whereas in the RFG model
$f(\psi^{\prime})=0$ for $\psi^{\prime} \leq -1$. It was shown in
\cite{AGK+05} that the QE scaling function can be obtained within
the CDFM in two equivalent ways, on the basis of the local density
distribution, as well as of the nucleon momentum distribution. As
pointed out in \cite{AGK+05}, the nucleon momentum distributions
$n(k)$ for various nuclei obtained in \cite{AGI+02} within a
parameter-free theoretical approach based on the light-front
dynamics method (e.g., \cite{CK95,CDK98} and references therein)
can also be used to describe both $y$- and $\psi^{\prime}$-scaling
data. So, in our present work we explore both methods, CDFM and
LFD, to investigate further the scaling functions and their
applications to analyses of electron- and neutrino scattering off
nuclei.

Our work is motivated by the fact that different models of the
nuclear dynamics (including those with RMF dynamics and with
RPA-type correlations accounted for) describe with different
success the basic size and shape of the cross sections in studies
of high-energy inclusive lepton scattering used so far. For this
reason we extend further our consideration and calculate within
the CDFM and LFD the scaling functions in the kinematical regions
of the QE and $\Delta$ peak on the basis of momentum and density
distributions of finite nuclear systems in which nucleon
correlations are included. This can be done either by using
available empirical data for these quantities or theoretical
calculations in which correlations are included to some extent.
Then, the obtained scaling functions are applied to calculate
electron-nucleus cross sections in QE and $\Delta$ regions in the
energy range from 500 MeV to 2 GeV for the target nucleus $^{12}$C
and to predict charge-changing neutrino and antineutrino reaction
cross sections from the scaling region to the QE peak at energies
of few GeV. We also make comparisons of the results obtained using
our methods with those obtained using the RFG model and other
theoretical schemes.

The paper is organized in the following way. In Sec. II we present
the formalism needed in studies of scaling functions in the
quasielastic region and validate the superscaling within the CDFM
and LFD for a variety of nuclei with $Z=N$ and $Z\neq N$. Then, we
consider the nucleon momentum distributions and their applications
in both approaches showing the sensitivity of the calculated
scaling functions to the peculiarities of $n(k)$ in different
regions of momenta. Section III contains the CDFM and LFD methods
to build up the scaling function in the $\Delta$ region. The
formalism involved in obtaining the electron-nucleus cross
sections in QE and $\Delta$ kinematical regions and the results of
the numerical calculations are presented in Sec. IV A. In Sec. IV
B we give our theoretical predictions for cross sections of
quasielastic charge-changing neutrino reactions. Finally, in Sec.
V we summarize the results of our work.

\section{Scaling function in the quasielastic region}
\subsection{QE scaling function in the CDFM}

As already mentioned in the Introduction, the superscaling
behavior was firstly considered within the framework of the RFG
model \cite{AMD+88,BCD+98,DS99l,DS99,MDS02,BCD+04} where a
properly defined function of the $\psi^{\prime}$-variable was
introduced. As pointed out in \cite{DS99}, however, the actual
nuclear dynamical content of the superscaling is more complex than
that provided by the RFG model. It was observed that the
experimental data have a superscaling behavior in the low-$\omega$
side ($\omega$ being the transfer energy) of the quasielastic peak
for large negative values of $\psi^{\prime}$ (up to
$\psi^{\prime}\approx -2$), while the predictions of the RFG model
are $f(\psi^{\prime})=0$ for $\psi^{\prime}\leq -1$. This imposes
the consideration of the superscaling in realistic finite systems.
One of the approaches to do this was developed
\cite{AGK+04,AGK+05} in the CDFM \cite{AHP88,AHP93,ANP79+,A+89+}
which is related to the $\delta$-function limit of the generator
coordinate method \cite{AGK+04,Grif57}. It was shown in
\cite{AGK+04,AGK+05,AIG+06} that the superscaling in nuclei can be
explained quantitatively on the basis of the similar behavior of
the high-momentum components of the nucleon momentum distribution
in light, medium and heavy nuclei. As already mentioned, the
latter is related to the effects of the NN correlations in nuclei
(see, e.g. \cite{AHP88,AHP93}).

The scaling function in the CDFM was obtained starting from that
in the RFG model \cite{AMD+88,BCD+98,DS99l,DS99} in two equivalent
ways, on the basis of the local density distribution $\rho(r)$ and
of the nucleon momentum distribution $n(k)$. This allows one to
study simultaneously the role of the NN correlations included in
$\rho(r)$ and $n(k)$ in the case of the superscaling phenomenon.
To explore these properties the scaling function
$f(\psi^{\prime})$ has been derived in two ways in CDFM in
\cite{AGK+05}. Firstly, by means of the density distribution
$\rho(r)$, it leads to
\begin{equation}
f^{QE}(\psi')= \int_{0}^{\alpha/(k_{F}|\psi'|)}dR |F(R)|^{2}
f_{RFG}^{QE}(\psi'(R)),
\label{eq:1}
\end{equation}
with a weight function of the form
\begin{equation}
|F(R)|^{2}=-\frac{1}{\rho_{0}(R)} \left. \frac{d\rho(r)}{dr}\right
|_{r=R},
\label{eq:2}
\end{equation}
where
\begin{equation}
\rho_{0}(R)=\frac{3A}{4\pi R^{3}}.
\label{eq:3}
\end{equation}
$f_{RFG}^{QE}(\psi^{\prime}(R))$ with
$\psi^{\prime}(R)=k_{F}R\psi^{\prime}/\alpha$ is the scaling
function related to the RFG model
\begin{eqnarray}
f_{RFG}^{QE}(\psi'(R))& =& \displaystyle \frac{3}{4} \left[
1-\left( \frac{k_FR|\psi'|}{\alpha} \right)^{2}\right] \left\{ 1+
\left( \frac{Rm_N}{\alpha}\right)^2 \left(
\frac{k_FR|\psi'|}{\alpha}
\right)^2 \right. \nonumber\\
&& \times \displaystyle \left. \left[2+ \left( \frac{\alpha}{Rm_N}
\right)^2- 2\sqrt{1+ \left( \frac{\alpha}{Rm_N} \right)^2}\right]
\right\},
\label{eq:4}
\end{eqnarray}
$m_{N}$ being the nucleon mass and $\alpha=(9\pi A/8)^{1/3}\simeq
1.52A^{1/3}$. Secondly, by means of the momentum distribution
$n(k)$, the scaling function is expressed by
\begin{equation}
f^{QE}(\psi')= \int_{k_{F}|\psi'|}^{\infty} d\overline{k}_F
|G(\overline{k}_F)|^2 f_{RFG}^{QE}(\psi'(\overline{k}_F)),
\label{eq:5}
\end{equation}
where $\psi'(\overline{k}_F)=k_{F}\psi'/\overline{k}_F$ and the
weight function is
\begin{equation}
|G(\overline{k}_F)|^2=- \frac{1}{n_0(\overline{k}_F)}\left.
\frac{dn(k)}{dk}\right |_{k=\overline{k}_F}
\label{eq:6}
\end{equation}
with
\begin{equation}
n_0(\overline{k}_F)= \frac{3A}{4\pi {\overline{k}_F}^3}.
\label{eq:7}
\end{equation}
In Eq.~(\ref{eq:5}) the RFG scaling function
$f_{RFG}^{QE}(\psi'(\overline{k}_F))$ can be obtained from
$f_{RFG}^{QE}(\psi'(R))$ [Eq.~(\ref{eq:4})] by changing there
$\alpha/R$ by $\overline{k}_F$. In Eqs.~(\ref{eq:1}), (\ref{eq:4})
and (\ref{eq:5}) the Fermi momentum $k_{F}$ is not a free
parameter for different nuclei as it is in the RFG model, but
$k_{F}$ is calculated within the CDFM for each nucleus using the
corresponding expressions:
\begin{equation}
k_F= \int_{0}^{\infty} dR k_{F}(R)|F(R)|^2= \alpha \int_{0}^{\infty}
dR \frac{1}{R}|F(R)|^{2}= \frac{4\pi(9\pi)^{1/3}}{3A^{2/3}}
\int_{0}^{\infty} dR \rho(R) R
\label{eq:8}
\end{equation}
when the condition
\begin{equation}
\lim_{R\rightarrow \infty} \left[ \rho(R)R^2 \right]=0
\label{eq:9}
\end{equation}
is fulfilled and
\begin{equation}
k_F= \frac{16\pi}{3A} \int_0^\infty d\overline{k}_F n(\overline{k}_F
) {\overline{k}_F}^3
\label{eq:10}
\end{equation}
when the condition
\begin{equation}
\lim_{\overline{k}_F\rightarrow \infty}\left[
n(\overline{k}_F){\overline{k}_F}^4 \right]=0
\label{eq:11}
\end{equation}
is fulfilled.

As shown in \cite{AGK+05}, the integration in Eqs.~(\ref{eq:1})
and (\ref{eq:5}), using Eqs.~(\ref{eq:2}) and (\ref{eq:6}), leads
to the explicit relationships of the scaling functions with the
density and momentum distributions:
\begin{equation}
f^{QE}(\psi')= \frac{4\pi}{A}\int_{0}^{\alpha/(k_{F}|\psi'|)}dR
\rho(R) \left[ R^2 f_{RFG}^{QE}(\psi'(R))+ \frac{R^3}{3}
\frac{\partial f_{RFG}^{QE}(\psi'(R))}{\partial R} \right]
\label{eq:12}
\end{equation}
and
\begin{equation}
f^{QE}(\psi')= \frac{4\pi}{A} \int_{k_{F}|\psi'|}^{\infty}
d\overline{k}_F n(\overline{k}_F) \left[ {\overline{k}_F}^2
f_{RFG}^{QE}(\psi'(\overline{k}_F))+ \frac{{\overline{k}_F}^3}{3}
\frac{\partial f_{RFG}^{QE}(\psi'(\overline{k}_F))}{\partial
\overline{k}_F} \right],
\label{eq:13}
\end{equation}
the latter at
\begin{equation}
\lim_{\overline{k}_F\rightarrow \infty}\left[
n(\overline{k}_F){\overline{k}_F}^3 \right]=0.
\label{eq:14}
\end{equation}
One can see the symmetry in both Eqs.~(\ref{eq:12}) and
(\ref{eq:13}) written in $r$- and $k$-space. We also note that in
the consideration up to here the CDFM scaling function
$f^{QE}(\psi^{\prime})$ is symmetric under the change of
$\psi^{\prime}$ by -$\psi^{\prime}$.

\begin{figure}[htb]
\includegraphics[width=10cm]{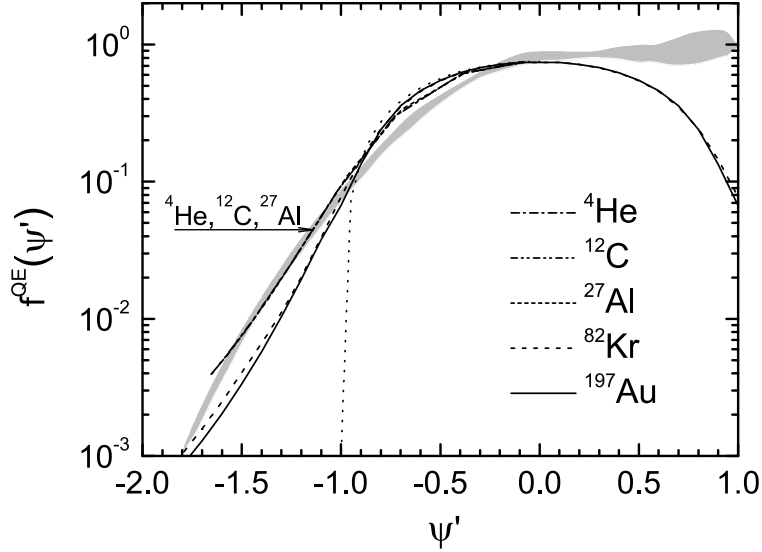}
\caption{The quasielastic scaling function $f^{QE}(\psi^{\prime})$
at $q=1000$ MeV/c for $^{4}$He, $^{12}$C, $^{27}$Al, $^{82}$Kr and
$^{197}$Au calculated in CDFM. Dotted line: RFG model result. The
curves for $^{4}$He, $^{12}$C and $^{27}$Al nuclei almost
coincide. Grey area: experimental data \protect\cite{DS99l,DS99}.
\label{fig01}}
\end{figure}

In Refs.~\cite{AGK+04,AGK+05} we used the charge density
distributions to determine the weight function $|F(R)|^{2}$ and
$f^{QE}(\psi^{\prime})$ in Eqs.~(\ref{eq:1}), (\ref{eq:2}) and
(\ref{eq:8}) for the cases of $^{4}$He, $^{12}$C, $^{27}$Al,
$^{56}$Fe and $^{197}$Au. The results for the scaling function
$f^{QE}(\psi^{\prime})$ agree well with the available data from
the inclusive quasielastic electron scattering for $^{4}$He,
$^{12}$C, $^{27}$Al, $^{56}$Fe and only approximately for
$^{197}$Au for various values of the transfer momentum $q=500,
1000, 1650$ MeV/c \cite{AGK+04} and $q=1560$ MeV/c \cite{AGK+05},
showing superscaling for negative values of $\psi^{\prime}$
including also those smaller than -1, whereas in the RFG model
$f(\psi^{\prime})=0$ for $\psi^{\prime} \leq -1$. One can see this
in Fig.~\ref{fig01} for $^{4}$He, $^{12}$C and $^{27}$Al at
$q=1000$ MeV/c. At the same time, however, in \cite{AGK+04,AGK+05}
we encountered some difficulties to describe the superscaling in
$^{197}$Au which was the heaviest nucleus considered. We related
this in \cite{AGK+04,AGK+05} to the particular A-dependence of
$n(k)$ in the model that does not lead to realistic high-momentum
components of $n(k)$ in the heaviest nuclei. We followed in
Refs.~\cite{AGK+04,AGK+05} an artificial way to "improve" the
high-momentum tail of $n(k)$ in $^{197}$Au by taking the value of
the diffuseness parameter $b$ in the Fermi-type charge density
distribution of this nucleus to be $b$=1 fm instead of the value
$b$=0.449 fm (as obtained from electron elastic scattering
experiments, see e.g. \cite{PP03}). In this way the high-momentum
tail of $n(k)$ for $^{197}$Au in CDFM becomes similar to those of
$^{4}$He, $^{12}$C, $^{27}$Al, and $^{56}$Fe and this leads to a
good agreement of the scaling function $f^{QE}(\psi^{\prime})$
with the data also for $^{197}$Au. Still in \cite{AGK+04} we
pointed out, however, that all the nucleons (not just the protons)
may contribute to $f^{QE}(\psi^{\prime})$ for the transverse
electron scattering and this could be simulated by increasing of
the diffuseness of the matter density with respect to that of the
charge density for a nucleus like $^{197}$Au that has much larger
number of neutron than of protons.

In \cite{AIG+06} we assumed that the reason why the CDFM does not
work properly in the case of $^{197}$Au is that we had used in
\cite{AGK+04,AGK+05} only the phenomenological charge density,
while this nucleus has many more neutrons than protons ($N$=118
and $Z$=79) and therefore, proton and neutron densities may differ
considerably. In the case when $Z\neq N$ and the proton and
neutron densities are not similar, the total scaling function will
be expressed by the sum of the proton $f_{p}^{QE}(\psi^{\prime})$
and neutron $f_{n}^{QE}(\psi^{\prime})$ scaling functions which
are determined by the proton and neutron densities $\rho_{p}(r)$
and $\rho_{n}(r)$, respectively:
\begin{equation}
f_{p(n)}^{QE}(\psi^{\prime})=\!\!\!\!\int\limits_{0}^{\alpha_{p(n)}/(k^{p(n)}_{F}
|\psi^{\prime}|)}\!\!\!\!dR
|F_{p(n)}(R)|^{2}f_{RFG}^{p(n)}(\psi'(R)).
\label{eq:15}
\end{equation}
In Eq.~(\ref{eq:15}) the proton and neutron weight functions are
obtained from the corresponding proton and neutron densities
\begin{equation}
\left|F_{p(n)}(R)\right|^2=-\dfrac{4\pi
R^3}{3Z(N)}\left.\dfrac{d\rho_{p(n)}(r)}{dr}\right|_{r=R},
\label{eq:16}
\end{equation}
\begin{equation}
\alpha_{p(n)}=\left(\dfrac{9\pi Z(N)}{4}\right)^{1/3},
\label{eq:17}
\end{equation}
\begin{equation}
\int\limits_{0}^{\infty}\rho_{p(n)}(\vec{r})d\vec{r}=Z(N),
\label{eq:18}
\end{equation}
and the Fermi momentum for the protons and neutrons is given by
\begin{equation}
k_{F}^{p(n)}=\alpha_{p(n)}\int\limits_{0}^{\infty}dR
\frac{1}{R}|{F}_{p(n)}(R)|^{2}.
\label{eq:19}
\end{equation}
The RFG proton and neutron scaling functions
$f_{RFG}^{p(n)}(\psi'(R))$ have the form of Eq.~(\ref{eq:4}),
where $\alpha$ and $k_{F}$ stand for $\alpha_{p(n)}$ from
Eq.~(\ref{eq:17}) and $k_{F}^{p(n)}$ from Eq.~(\ref{eq:19}),
respectively. The functions are normalized as follows:
\begin{equation}
\int\limits_{0}^{\infty}|F_{p(n)}(R)|^{2}dR=1,
\label{eq:20}
\end{equation}
\begin{equation}
\int\limits_{-\infty}^{\infty}f_{p(n)}^{QE}(\psi^{\prime})d\psi^{\prime}=1.
\label{eq:21}
\end{equation}
Then the total scaling function can be expressed by means of both
proton and neutron scaling functions:
\begin{equation}
f^{QE}(\psi^{\prime})=\dfrac{1}{A}[Zf_p^{QE}(\psi^{\prime})+Nf_n^{QE}(\psi^{\prime})]
\label{eq:22}
\end{equation}
and is normalized to unity.

The same consideration can be performed equivalently on the basis
of the nucleon momentum distributions for protons $n^{p}(k)$ and
neutrons $n^{n}(k)$ presenting $f^{QE}(\psi^{\prime})$ by the sum
of proton and neutron scaling functions (\ref{eq:22}) calculated
similarly to Eqs.~(\ref{eq:15})-(\ref{eq:22}) (and to
Eqs.~(\ref{eq:5}), (\ref{eq:6}), (\ref{eq:10}) and (\ref{eq:11})):
\begin{equation}
f_{p(n)}^{QE}(\psi')=\!\!
\int\limits_{k^{p(n)}_{F}|\psi'|}^{\infty}\!\! d\overline{k}_F
|G_{p(n)}(\overline{k}_F)|^2
f^{p(n)}_{RFG}(\psi'(\overline{k}_F)),
\label{eq:23}
\end{equation}
where
\begin{equation}
|G_{p(n)}(\overline{k}_F)|^2=- \frac{4\pi
{\overline{k}_F}^3}{3Z(N)}\left. \frac{dn^{p(n)}(k)}{dk}\right
|_{k=\overline{k}_F}
\label{eq:24}
\end{equation}
with $f_{RFG}^{p(n)}(\psi'(\overline{k}_F))$ containing
$\alpha_{p(n)}$ from Eq.~(\ref{eq:17}) and $k_F^{p(n)}$ calculated
as
\begin{equation}
k^{p(n)}_F=  \int\limits_0^\infty d\overline{k}_F \overline{k}_F
|G_{p(n)}(\overline{k}_F)|^2.
\label{eq:25}
\end{equation}
The scaling functions for several examples, such as the medium
stable nuclei $^{62}$Ni and $^{82}$Kr and the heavy nuclei
$^{118}$Sn and $^{197}$Au are calculated following
Eqs.~(\ref{eq:15})-(\ref{eq:22}) using the corresponding proton
and neutron densities obtained in deformed self-consistent
mean-field (HF+BCS) calculations with density-dependent Skyrme
effective interaction (SG2) and a large harmonic-oscillator basis
with 11 major shells~\cite{Sarriguren99,Vautherin73}. In
Fig.~\ref{fig01} we give the results for the $^{82}$Kr and
$^{197}$Au nuclei in which $Z\neq N$ and compare them with the
results for $^{4}$He, $^{12}$C ($Z=N$) and $^{27}$Al ($Z\simeq
N$), as well as with the experimental data (presented by a grey
area and taken from \cite{DS99}) obtained for $^{4}$He, $^{12}$C,
$^{27}$Al, $^{56}$Fe, and $^{197}$Au. The scaling functions are in
a reasonable agreement with the data, which was not the case for
$^{197}$Au calculated in \cite{AGK+04} by using only the
Fermi-type charge density with phenomenological parameter values
$b=0.449$ fm and $R=6.419$ fm from \cite{PP03}. At the same time
we note also the improvement in comparison with the RFG model
result in which $f^{QE}(\psi^{\prime})=0$ for $\psi^{\prime}\leq
-1$. Thus, it can be concluded that the scaling function
$f^{QE}(\psi^{\prime})$ for nuclei with $Z\neq N$ for which the
proton and neutron densities are not similar has to be expressed
by the sum of the proton and neutron scaling functions. The latter
can be calculated by means of theoretically and/or experimentally
obtained proton and neutron local density distributions or
momentum distributions.

As known (e.g. \cite{Amaro2005,DS99}), the total inclusive
electron scattering response is assumed to be composed of several
contributions: i) the entire longitudinal contribution which
superscales and is represented by the QE scaling function
$f^{QE}(\psi^{\prime})$; ii) a part of the transverse response,
which arises from the quasielastic knockout of nucleons and is
also driven by the scaling function $f^{QE}(\psi^{\prime})$, and
iii) the additional contribution of the transverse response from
MEC effects and from inelastic single-nucleon processes including
the excitation of the $\Delta$ isobar. The effects of point iii)
break the scaling. In \cite{Amaro2005} an universal scaling
function $f^{QE}(\psi^{\prime})$ has been determined by reliable
separations of the empirical data into their longitudinal and
transverse contributions for $A>4$. Such separations are available
only for a few nuclei \cite{Jordan96}. All of these response
functions have been used to extract the "universal" QE response
function $f^{QE}(\psi^{\prime})$ (see Fig.~1 of \cite{Amaro2005})
which is parametrized by a simple function. This function has a
somewhat asymmetric shape. Its left tail $(\psi^{\prime}<0)$
passes through the grey area of Fig.~\ref{fig01}. The right tail
$(\psi^{\prime}>0)$ extends larger towards positive values of
$\psi^{\prime}$. In contrast, the RFG scaling function is
symmetric. The CDFM scaling function discussed so far, which is
based on the RFG one, is also symmetric. As mentioned, the maximum
value of $f^{QE}(\psi^{\prime})$ in RFG (and in CDFM) is 3/4,
while the empirical scaling function extracted in
Ref.~\cite{Amaro2005} reaches about 0.6.

\begin{figure}[htb]
\includegraphics[width=10cm]{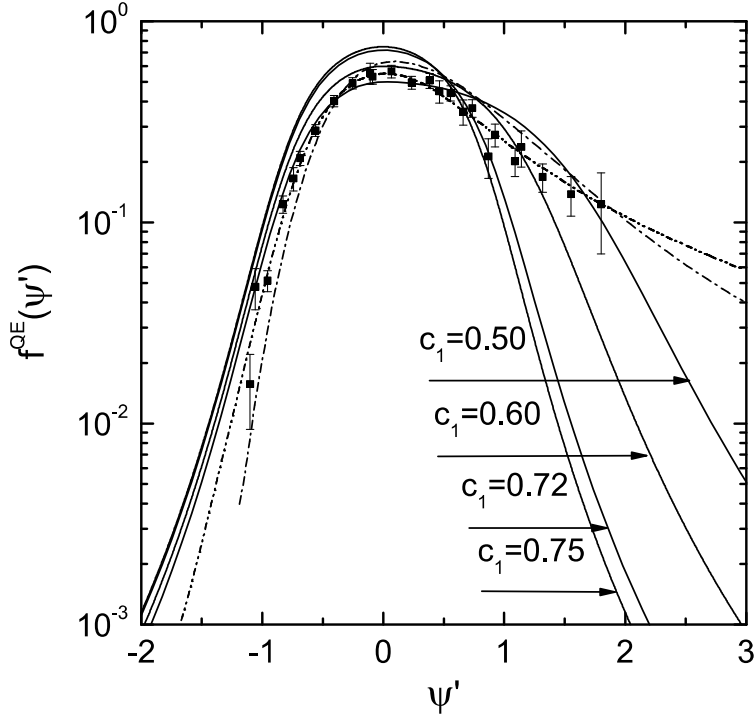}
\caption{The quasielastic scaling function $f^{QE}(\psi^{\prime})$
for $^{12}$C calculated in CDFM in comparison with the
experimental data (black squares) \protect\cite{Amaro2005}. The
CDFM results for different values of $c_{1}$ are presented by
solid lines. Also shown for comparison is the phenomenological
curve which fits the data (dash-two dots), as well as the curve
(dash-dot line) corresponding to the result for $(e,e^{\prime})$
obtained within the relativistic impulse approximation and FSI
using the relativistic mean field (see
Refs.~\protect\cite{Caballero2005,Caballero2006}).
\label{fig02}}
\end{figure}

As mentioned in \cite{Amaro2005}, if FSI are neglected, the RMF
theory \cite{Kim95,Alberico97,Maieron2003} and relativized
shell-model studies \cite{Amaro96} provide rather modest
differences from the RFG predictions. Another possible reason for
the differences between the RFG (or mean-field results) and the
empirically determined scaling function arises from high-momentum
components in realistic wave functions which may be large enough.
In \cite{Amaro2005} the scaling function was taken from the
experiment. In the present work we also limit our approach to
phenomenology when considering the asymmetric shape and the
maximum value of the quasielastic scaling function. In order to
simulate the role of all the effects which lead to asymmetry, we
impose the latter on the RFG scaling function (and,
correspondingly, on the CDFM one) by introducing a parameter which
gives the correct maximum value of the scaling function ($c_{1}$
in our expressions given below) and also an asymmetry in
$f^{QE}(\psi^{\prime})$ for $\psi^{\prime}\geq 0$. We consider the
main parts of the RFG scaling function for $\psi^{\prime}\leq 0$
and $\psi^{\prime}\geq 0$ in the following forms, keeping the
parabolic dependence on $\psi^{\prime}$  as required in
Ref.~\cite{AMD+88}:
\begin{equation}
f_{RFG,1}^{QE}(\psi^{\prime})=c_{1}(1-\psi^{\prime
2})\Theta(1-\psi^{\prime 2}), \;\;\;\;\; \psi^{\prime}\leq 0,
\label{eq:26}
\end{equation}
\begin{equation}
f_{RFG,2}^{QE}(\psi^{\prime})=c_{1}\left[1-\left(\frac{\psi^{\prime}}{c_{2}}\right
)^{2}\right ]\Theta\left[1-\left(\frac{\psi^{\prime}}{c_{2}}\right
)^{2}\right ], \;\;\;\;\; \psi^{\prime}\geq 0.
\label{eq:27}
\end{equation}
The total RFG scaling function is normalized to unity:
\begin{equation}
\int_{-\infty}^{\infty}
f^{QE}_{RFG}(\psi^{\prime})d\psi^{\prime}=\int_{-\infty}^{\infty}
[f_{RFG,1}^{QE}(\psi^{\prime})+f_{RFG,2}^{QE}(\psi^{\prime})]
d\psi^{\prime}=1.
\label{eq:28}
\end{equation}
If the normalization of the scaling function for negative values of
$\psi^{\prime}$ is equal to
\begin{equation}
a=\int_{-\infty}^{0} d\psi^{\prime}
f_{RFG,1}^{QE}(\psi^{\prime})=\frac{2}{3}c_{1},
\label{eq:29}
\end{equation}
then, in order to keep the total normalization [Eq.~(\ref{eq:28})],
the normalization for positive $\psi^{\prime}$ has to be:
\begin{equation}
1-a=\int_{0}^{\infty} d\psi^{\prime}
f_{RFG,2}^{QE}(\psi^{\prime})=\frac{2}{3}c_{1}c_{2}.
\label{eq:30}
\end{equation}
From Eqs.~(\ref{eq:29}) and (\ref{eq:30}) we get the relationship
between $c_{2}$ and $c_{1}$:
\begin{equation}
c_{2}=\frac{3}{2c_{1}}-1.
\label{eq:31}
\end{equation}
In the RFG $c_{1}=3/4$ and, correspondingly, $c_{2}=1$. In the CDFM
the QE scaling function will be:
\begin{equation}
f^{QE}(\psi^{\prime})=f_{1}^{QE}(\psi^{\prime})+f_{2}^{QE}(\psi^{\prime}),
\label{eq:32}
\end{equation}
where
\begin{equation}
f_{1}^{QE}(\psi^{\prime})\cong
\int_{0}^{\alpha/k_{F}|\psi^{\prime}|} dR |F(R)|^{2}
c_{1}\left[1-\left(\frac{k_{F}R|\psi^{\prime}|}{\alpha}\right
)^{2}\right ], \psi^{\prime}\leq 0,
\label{eq:33}
\end{equation}
\begin{equation}
f_{2}^{QE}(\psi^{\prime})\cong
\int_{0}^{c_{2}\alpha/k_{F}|\psi^{\prime}|} dR |F(R)|^{2}
c_{1}\left[1-\left(\frac{k_{F}R|\psi^{\prime}|}{c_{2}\alpha}\right
)^{2}\right ], \psi^{\prime}\geq 0.
\label{eq:34}
\end{equation}
In this approach, parametrizing the RFG scaling function by the
coefficient $c_{1}$ we account for the experimental fact that
$c_{1}\neq 3/4$ and take this value in accordance with the
empirical data. Then from the normalization
[Eqs.~(\ref{eq:28})-(\ref{eq:30})] we determine the corresponding
value of $c_{2}$ using Eq.~(\ref{eq:31}). As in
\cite{AGK+04,AGK+05}, the CDFM scaling function is obtained
[Eqs.~(\ref{eq:32})-(\ref{eq:34})] by averaging the RFG scaling
function. As an example, we give in Fig.~\ref{fig02} the CDFM QE
scaling function for different values of $c_{1}$ (0.75, 0.72, 0.60
and 0.50) in comparison with the empirical data and the
phenomenological fit. We also include for reference the scaling
function obtained from calculations for $(e,e^{\prime})$ reaction
based on the relativistic impulse approximation (RIA) with FSI
described using the RMF potential
(see~\cite{Caballero2005,Caballero2006} for details). In this way
we simulate in a phenomenological way the role of the effects
which violate the symmetry for positive values of $\psi^{\prime}$
of the QE scaling function, which in the RMF approximation are
seen to be due to the FSI.

\subsection{QE scaling function in the LFD method}

In this Subsection we will obtain the QE scaling function on the
basis of calculations of nucleon momentum distribution (using
Eqs.~(5)--(7) or Eq.~(13)) obtained within a modification of the
approach from~\cite{AGI+02}. The latter is based on the momentum
distribution in the deuteron from the light-front dynamics (LFD)
method (e.g., \cite{CK95,CDK98} and references therein). Using the
natural-orbital representation of the one-body density
matrix~\cite{Low55}, $n(k)$ was written as a sum of contributions
from hole-states $[n^{\text{h}}(k)]$ and particle-states
$[n^{\text{p}}(k)]$ (see also~\cite{AGK+05})
\begin{equation}
n_A(k)=N_A\left[ n^{\text{h}}(k)+ n^{\text{p}}(k)\right] .
\label{eq:35}
\end{equation}
In~(\ref{eq:35})
\begin{equation}
n^{\text{h}}(k)= C(k)\sum_{nlj}^{\text{F.L.}} 2(2j+1)
\lambda_{nlj}
 |R_{nlj}(k)|^2 ,
\label{eq:36}
\end{equation}
where F.L. denotes the Fermi level, and
\begin{equation}
C(k)= \frac{m_N}{(2\pi)^3\sqrt{k^2+m_N^2}} ,
\label{eq:37}
\end{equation}
$m_N$ being the nucleon mass. To a good approximation for the hole
states, the natural occupation numbers $\lambda_{nlj}$ are close
to unity in~\cite{AGI+02} and the natural orbitals $R_{nlj}(k)$
are replaced by single-particle wave functions from the
self-consistent mean-field approximation. In~\cite{AGI+02}
Woods-Saxon single-particle wave functions were used for protons
and neutrons. $N_A$ is the normalization factor. Concerning  the
particle-state $[n^{\text{p}}(k)]$ contribution in~(\ref{eq:35}),
we used in~\cite{AGI+02} and~\cite{AGK+05} the well-known facts
that: (i) the high-momentum components of $n(k)$ caused by
short-range and tensor correlations are almost completely
determined by the contributions of the particle-state natural
orbitals (e.g.~\cite{SAD93}), and (ii) the high-momentum tails of
$n_{A}(k)/A$ are approximately equal for all nuclei and are a
rescaled version of the nucleon momentum distribution in the
deuteron $n_d(k)$~\cite{FCW00},
\begin{equation}
n_A(k)\simeq \alpha_A n_d(k),
\label{eq:38}
\end{equation}
where $\alpha_A$ is a constant. These facts made it possible to
assume in~\cite{AGI+02} and~\cite{AGK+05} that $n^{\text{p}}(k)$
is related to the high-momentum components $n_5(k)$ of the
deuteron, that is,
\begin{equation}
n^{\text{p}}(k)= \frac{A}{2} n_5(k).
\label{eq:39}
\end{equation}
In~(\ref{eq:39}) $n_5(k)$ is expressed by an angle-averaged
function~\cite{AGI+02} as
\begin{equation}
n_5(k)= C(k) \overline{(1-z^2) f_5^2(k)}.
\label{eq:40}
\end{equation}
In~(\ref{eq:40}) $z=\cos(\widehat{\vec{k},\vec{n}})$, $\vec{n}$
being a unit vector along the three vector ($\vec{\omega}$)
component of the four-vector $\omega$ which determines the
position of the light-front surface~\cite{CK95,CDK98}. The
function $f_5(k)$ is one of the six scalar functions
$f_{1-6}(k^2,\vec{n}\cdot\vec{k})$ which are the components of the
deuteron total wave function $\Psi(\vec{k},\vec{n})$. It was
shown~\cite{CK95} that $f_5$ largely exceeds other $f$-components
for $k\geq$ 2.0--2.5 fm$^{-1}$ and is the main contribution to the
high-momentum component of $n_d(k)$, incorporating the main part
of the short-range properties of the nucleon-nucleon interaction.

\begin{figure}[htb]
\includegraphics[width=10cm]{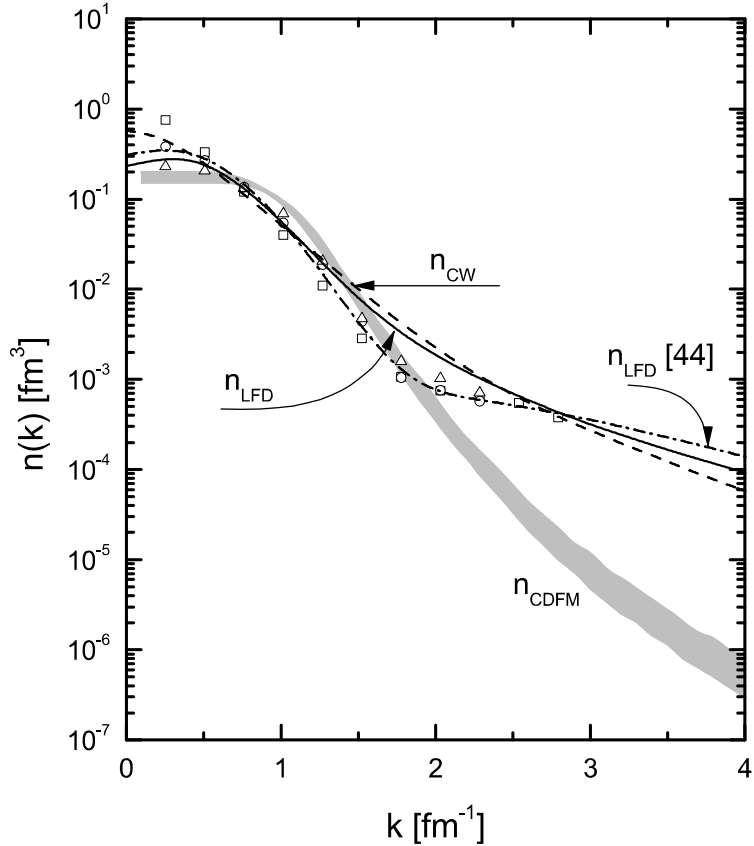}
\caption{The nucleon momentum distribution $n(k)$. Grey area: CDFM
combined results for $^{4}$He, $^{12}$C, $^{27}$Al, $^{56}$Fe and
$^{197}$Au. Solid line: result of the present work for $^{12}$C
using the modified LFD approach with $\beta=0.80$. Dashed line:
$y_{CW}$-scaling result \protect\cite{CW99,CW97}. Dash-dotted
line: result of LFD for $^{12}$C from \protect\cite{AGI+02}.
Dotted line: the mean-field result using Wood-Saxon
single-particle wave functions for $^{56}$Fe. Open squares,
circles and triangles are $y$-scaling data \protect\cite{Cio90}
for $^{4}$He, $^{12}$C and $^{56}$Fe, respectively. The
normalization is: $\int n(k)d^{3}{\bf k}=1$.
\label{fig03}}
\end{figure}

\begin{figure}[htb]
\includegraphics[width=10cm]{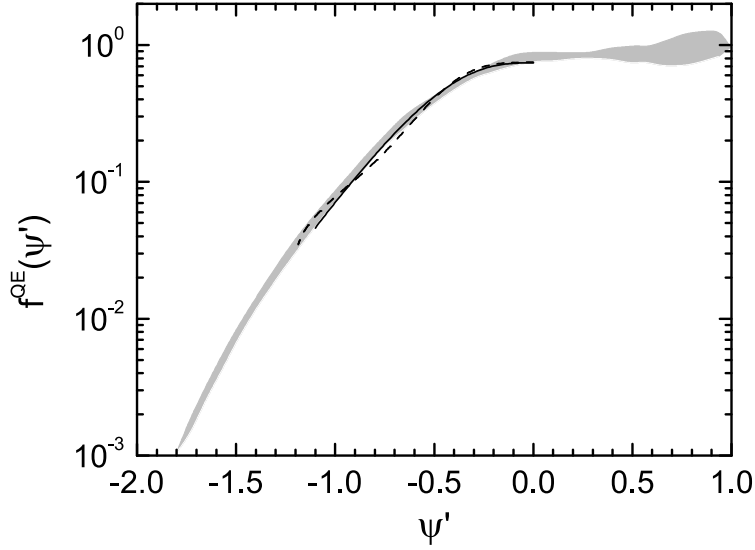}
\caption{The quasielastic scaling function $f^{QE}(\psi^{\prime})$
calculated using Eq.~(\ref{eq:41}) at $q=1000$ MeV/c with
$n_{CW}(k)$ from the $y_{CW}$-scaling analysis
\protect\cite{CW99,CW97} for $^{56}$Fe (solid line) and
$n_{LFD}(k)$ from modified LFD approach [Eq.~(\ref{eq:44})] for
$^{12}$C (dashed line). \label{fig04}}
\end{figure}

It was shown in Fig.~2 of~\cite{AGK+05} that the calculated LFD
$n(k)$'s are in good agreement with the ``$y$-scaling data'' for
$^4$He, $^{12}$C and $^{56}$Fe from~\cite{Cio90} and also with the
$y_{\text{CW}}$ analysis~\cite{CW99,CW97} up to $k\lesssim
2.8$~fm$^{-1}$. For  larger $k$ the momentum distributions from
LFD exceeds that obtained from $y_{\text{CW}}$ analysis. We should
note also that the calculated scaling function $f^{QE}(\psi')$
using the approximate relationship (see Eq.~(75) and Fig.~4
of~\cite{AGK+05})
\begin{equation}
f^{QE}(\psi')\simeq 3\pi \int_{|y|}^{\infty} d\overline{k}_F \,
n(\overline{k}_F) {\overline{k}_F}^2, \;\;\;
|y|=\frac{1-\sqrt{1-4ck_F |\psi'|}}{2c}, \;\;\; c\equiv
\frac{\sqrt{1+m_{N}^{2}/q^{2}}}{2m_{N}},
\label{eq:41}
\end{equation}
for $^{56}$Fe at $q=1000$~MeV/c is in agreement with the data for
$-0.5\lesssim\psi'\leq0$, while in the region
$-1.1\leq\psi'\leq-0.5$ it shows a dip in the interval
$-0.9\leq\psi'\leq-0.6$. This difference is due to the particular
form of $n(k)$  from LFD shown in Fig.~2 of \cite{AGK+05} (a dip
around $k\approx1.7$~fm$^{-1}$ and a very high-momentum tail at
$k\gtrsim2.8$~fm$^{-1}$). This result showed that the
assumption~(\ref{eq:39}) for the particle-state contribution is a
rather rough one. In this paper we consider a modification of the
approach in which we include partially in the particle-state part
$n^{p}(k)$ not only $n_5(k)$ but also $n_2(k)$ which is related to
the angle-averaged function $f_2(k)$:
\begin{equation}
n_2(k)= C(k) \overline{f_2^2(k)}.
\label{eq:42}
\end{equation}
Then the particle-state part can be written in the form
\begin{equation}
n^{\text{p}}(k)=\beta\big[n_2(k)+n_5(k)\big],
\label{eq:43}
\end{equation}
where $\beta $ is a parameter. Then the LFD nucleon momentum
distribution for the nucleus with $A$ nucleons will be:
\begin{equation}
n_{\text{LFD}}(k)=N_A\big[n^{\text{h}}(k)+\beta\big(n_2(k)+n_5(k)\big)\big],
\label{eq:44}
\end{equation}
with $n^{\text{h}}(k)$ from Eq.~(\ref{eq:36}) and
\begin{equation}
N_A= \left\{ 4\pi \int_0^\infty dq \, q^2 \left[
\sum_{nlj}^{\text{F.L.}} 2(2j+1) \lambda_{nlj} C(q) |R_{nlj}(q)|^2
+ \beta \big(n_2(q)+n_5(q)\big)\right] \right\}^{-1} .
\label{eq:45}
\end{equation}
In Fig.~\ref{fig03} we present the nucleon momentum distribution
for $^{12}$C calculated within the LFD method using
Eqs.~(\ref{eq:35})--(\ref{eq:37}), (\ref{eq:40}),
(\ref{eq:42})--(\ref{eq:45}) with the parameter value
$\beta=0.80$. It is compared with the band of CDFM momentum
distributions for $^4$He, $^{12}$C, $^{27}$Al, $^{56}$Fe,
$^{197}$Au (grey area), with $n_{\text{CW}}(k)$ from the
$y_{\text{CW}}$ analysis~\cite{CW99,CW97} and with the $y$-scaling
data~\cite{Cio90} for $^4$He, $^{12}$C, and $^{56}$Fe. It can be
seen that up to $k\simeq2.8$~fm$^{-1}$ $n_{\text{LFD}}$ curve is
close to the results of~\cite{Cio90,CW99,CW97}. For the region
$1\leq k\leq2.5$~fm$^{-1}$ it is between them and for
$k\geqslant2.8$~fm$^{-1}$ it is close to $n_{\text{CW}}(k)$, in
contrast to our previous results in Fig.~2 of Ref.~\cite{AGK+05}
(see also \cite{AGI+02}) which were based on Eq.~(\ref{eq:39}) and
which are also shown for comparison in Fig.~\ref{fig03}. This
behavior of $n_{\text{LFD}}(k)$ reflects in the result of the
calculation of the QE scaling function using Eq.~(\ref{eq:41})
given in Fig.~\ref{fig04}. It can be seen that both momentum
distributions $n_{\text{CW}}$~\cite{CW97} and $n_{\text{LFD}}(k)$
[Eq.~(\ref{eq:44})] give a good agreement with the experimental
data for the QE scaling function at least up to $\psi'\simeq-1.2$.
This result is an improvement of that for LFD shown in Fig.~4
of~\cite{AGK+05}, where only the contribution $n_5$ was used in
the calculation of $n^{\text{p}}(k)$~(\ref{eq:39}) and
$n_{\text{LFD}}(k)$.

\section{Scaling function in the quasielastic delta region}

In this Section we will extend our analysis within both CDFM and
LFD to the $\Delta$-peak region, which is not too far above the QE
peak region and is the main contribution to the inelastic
scattering. Dividing the cross section by the appropriate
single-nucleon cross section, now for $N\rightarrow\Delta$
transition, and displaying the results versus a new scaling
variable ($\psi_\Delta'$) (in which the kinematics of resonance
electro-production is accounted for) it is obtained in
~\cite{Amaro2005} that the results scale quite well. This is
considered as an indication that the procedure has identified the
dominant contributions not only in the QE region, but also in the
$\Delta$-region.

The shifted dimensionless scaling variable in the $\Delta$-region
$\psi_\Delta'$ is introduced (see, e.g.~\cite{Amaro2005}) by the
expression:
\begin{equation} \label{eq:46}
\psi _{\Delta}'\equiv \left[ \frac{1}{\xi _{F}}\left( \kappa
\sqrt{{{\rho}_{\Delta}'}^{2}+1/\tau' }-\lambda'
{\rho}_{\Delta}'-1\right) \right] ^{1/2}\times \left\{
\begin{array}{cc}
+1, & \lambda' \geq {\lambda'}_{\Delta}^{0} \\
-1, & \lambda' \leq {\lambda'}_{\Delta}^{0}
\end{array}
\right. ,
\end{equation}
where
\begin{equation}\label{eq:47}
\xi_F\equiv \sqrt{1+\eta_F^2}-1,\qquad \eta_F \equiv
\dfrac{k_F}{m_N},
\end{equation}
\begin{equation}\label{eq:48}
\lambda'=
\lambda-\dfrac{E_{shift}}{2m_N},\qquad\tau'=\kappa^2-\lambda'^2,
\end{equation}
\begin{equation}\label{eq:49}
\lambda =\dfrac{\omega}{2m_{N}},\qquad\kappa =
\dfrac{q}{2m_{N}},\qquad\tau =\kappa ^{2}-\lambda ^{2},
\end{equation}
\begin{equation}\label{eq:50}
{\lambda'}_{\Delta}^{0}=\lambda
_{\Delta}^{0}-\frac{E_{shift}}{2m_N},\qquad \lambda
_{\Delta}^{0}=\frac{1}{2}\left[ \sqrt{\mu _{\Delta}^{2}+4\kappa
^{2}}-1\right] ,
\end{equation}
\begin{equation}\label{eq:51}
\qquad \mu _{\Delta}=m_{\Delta }/m_{N},
\end{equation}
\begin{equation}\label{eq:52}
\rho_{\Delta} =1+\dfrac{\beta _{\Delta}}{\tau},\qquad
{\rho}_{\Delta}' =1+\dfrac{\beta _{\Delta}}{\tau'},
\end{equation}
\begin{equation}\label{eq:53}
\beta _{\Delta} =\dfrac{1}{4}\left( \mu _{\Delta}^{2}-1\right).
\end{equation}

The relativistic Fermi gas superscaling function in the $\Delta$
domain is given by~\cite{Amaro2005}:
\begin{equation}\label{eq:54}
f_{RFG}^{\Delta}(\psi'_{\Delta})=\dfrac{3}{4}(1-{\psi'_{\Delta}}^2)\Theta(1-
{\psi'_{\Delta}}^2).
\end{equation}

Following the CDFM application to the scaling phenomenon, the
$\Delta$-scaling function in the model will be:
\begin{equation}\label{eq:55}
f^{\Delta}(\psi'_{\Delta})=\int_{0}^{\infty}dR|F_\Delta(R)|^2f_{RFG}^{\Delta}(\psi'_{\Delta}(R)).
\end{equation}
In Eq.~(\ref{eq:55}):
\begin{equation}\label{eq:56}
{\psi'_{\Delta}}^2(R)=\dfrac{1}{\left[\sqrt{1+\dfrac{k_F^2(R)}{m_N^2}}-1\right]}
\left[  \kappa \sqrt{{{\rho}_{\Delta}'}^{2}+\dfrac{1}{\tau'}
}-\lambda' {\rho}_{\Delta}'-1\right]\equiv t(R).{\psi_\Delta'}^2,
\end{equation}
where
\begin{equation}\label{eq:57}
t(R)\equiv\dfrac{\left[\sqrt{1+\dfrac{k_F^2}{m_N^2}}-1\right]}{\left[\sqrt{1+\dfrac{k_F^2(R)}
{m_N^2}}-1\right]}\quad \text{and}\quad k_F(R)=\dfrac{\alpha}{R}.
\end{equation}

\begin{figure}[htb]
\includegraphics[width=10cm]{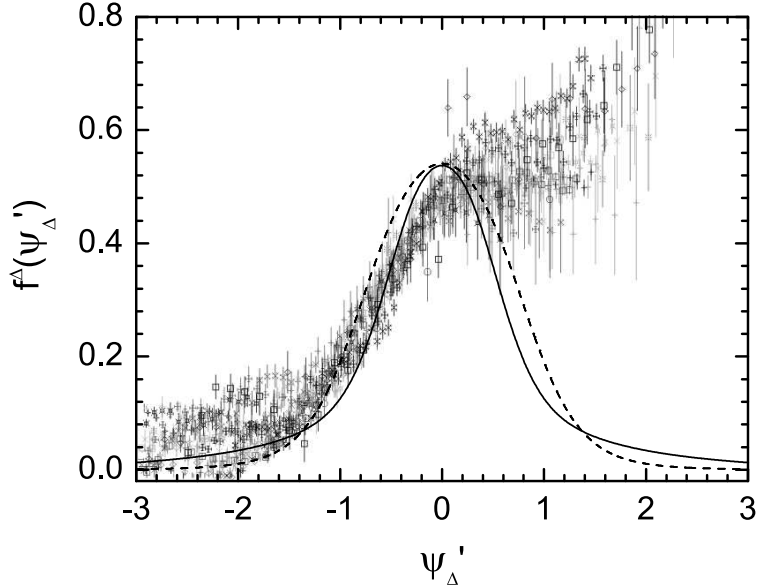}
\caption{The $f^{\Delta}(\psi_{\Delta}^{\prime})$ scaling function
for $^{12}$C in the $\Delta$-region. Dashed line: CDFM result
(with $R_{\Delta}=1.565$ fm, $b_{\Delta}=0.420$ fm, $k_{F}=1.20$
fm$^{-1}$). Solid line: result of modified LFD approach
($\beta=0.80$, $k_{F}=1.20$ fm$^{-1}$). The coefficient
$c_{1}=0.54$ in both CDFM and LFD cases. Averaged experimental
values of $f^{\Delta}(\psi_{\Delta}^{\prime})$ are taken from
\protect\cite{Amaro2005}.
\label{fig05}}
\end{figure}

In the CDFM $k_F$ can be calculated using the density distribution
(Eqs.~(\ref{eq:8}), (\ref{eq:9}) or (\ref{eq:19}) and
(\ref{eq:16})) or the momentum distribution (Eqs.~(\ref{eq:10}),
(\ref{eq:11}) or (\ref{eq:25}) and (\ref{eq:24})). The weight
function $|F_{\Delta}(R)|^2$ is related to the density
distributions (Eqs.~(\ref{eq:2}) or (\ref{eq:16})). In the
equivalent form of the CDFM, the scaling function can be written
in the form:
\begin{equation}\label{eq:58}
f^{\Delta}(\psi'_{\Delta})=\int_{0}^{\infty} d\overline{k}_F
|G_\Delta(\overline{k}_F)|^2
f_{RFG}^{\Delta}(\psi_\Delta'(\overline{k}_F)),
\end{equation}
where $G_\Delta(\overline{k}_F)$  is determined by means of the
momentum distribution (Eqs.~(\ref{eq:6}) or (\ref{eq:24})) and
\begin{equation}\label{eq:59}
\psi_\Delta'^{2}(\overline{k}_F)\equiv\widetilde{t}(\overline{k}_F).\psi_\Delta'^{2}
\end{equation}
with
\begin{equation}\label{eq:60a}
\widetilde{t}(\overline{k}_F)\equiv
\dfrac{\left[\sqrt{1+\dfrac{k_F^2}{m_N^2}}-1\right]}
{\left[\sqrt{1+\dfrac{\overline{k}_F^2}{m_N^2}}-1\right]}.
\end{equation}
Here we would like to note that though the functional forms of
$f^{\Delta}(\psi'_\Delta)$ [Eq.~(\ref{eq:55})] and the weight
function $|F_\Delta(R)|^2$ (Eqs.~(\ref{eq:2}) or (\ref{eq:16}))
are like before, i.e. as in the case of the QE region, the
parameters of the densities (e.g. the half-radius $R_\Delta$ and
the diffuseness $b_\Delta$ when Fermi-type forms have been used)
may be different from those ($R$ and $b$) in the QE case. Along
this line, we calculated firstly the scaling function
$f^{\Delta}(\psi_\Delta')$ by means of
Eqs.~(\ref{eq:55})--(\ref{eq:57}) using the Fermi-type density for
$^{12}$C. We found the values of $R_\Delta$ and $b_\Delta$ fitting
the scaling data at the $\Delta$ peak extracted from the
high-quality world data for inclusive electron scattering (given
in Fig.~2 of~\cite{Amaro2005}). Our results are presented in
Fig.~\ref{fig05}. As mentioned already in the QE case, the
empirical data require to use a value of the coefficient in the
right-hand side of Eq.~(\ref{eq:54}) for the RFG scaling functions
$f_{RFG}^{\Delta}(\psi_\Delta')$ different from $3/4$. In our
calculations in the $\Delta$-region we use the value $0.54$. We
found that reasonable agreement with the data can be achieved
using the parameter values $R_\Delta=1.565$~fm and
$b_\Delta=0.420$~fm (the Fermi-momentum value is taken to be
$k_F=1.20$~fm$^{-1}$ and this choice leads to normalization to
unity of $f_{RFG}^{\Delta}(\psi_\Delta')$). The value of
$R_\Delta$ is smaller than that used in the description of the QE
superscaling function for $^{12}$C~\cite{AGK+04,AGK+05}
($R=2.470$~fm) while the value of $b_\Delta$ is the same as $b$ in
the QE case. Secondly, we calculated $f^{\Delta}(\psi_\Delta')$
using Eqs.~(\ref{eq:58}), (\ref{eq:59}) and (\ref{eq:60a}). In
Eq.~(\ref{eq:58}) the weight function
$|G_\Delta(\overline{k}_F)|^2$ was determined by means of
Eq.~(\ref{eq:6}) and the nucleon momentum distribution
$n_\text{LFD}$ (Eqs.~(\ref{eq:44}) and (\ref{eq:45})) calculated
with the parameter value $\beta=0.80$ (shown in Fig.~\ref{fig03}).
We note that the use of $n_\text{LFD}(k)$  with this value of
$\beta$ gives simultaneously a reasonable agreement both with the
results for the momentum distribution from the $y$-scaling data
shown in Fig.~\ref{fig03}, as well as with the QE scaling function
shown in Fig.~\ref{fig04}.

\section{Scaling functions and inclusive lepton scattering}
\subsection{Scaling functions and $(e,e^{\prime})$ reaction cross sections}

In the beginning of this Subsection we will give some basic
relationships concerning inclusive electron scattering from
nuclei. An electron with four-momentum $k^{\mu}=(\epsilon,{\bf
k})$ is scattered through an angle $\theta$ to four-momentum
$k^{\prime\mu}=(\epsilon^{\prime},{\bf k^{\prime}})$. The
four-momentum transfer is then
\begin{equation}
Q^{\mu}=(k-k^{\prime})^{\mu}=(\omega,{\bf q}),
\label{eq:60}
\end{equation}
where $\omega=\epsilon-\epsilon^{\prime}$, $q=|{\bf q}|=|{\bf
k}-{\bf k^{\prime}}|$ and
\begin{equation}
Q^{2}=\omega^{2}-q^{2}\leq 0.
\label{eq:61}
\end{equation}

In the one-photon-exchange approximation, the double-differential
cross section in the laboratory system can be written in the form
(e.g. \cite{AMD+88}):
\begin{equation}
\frac{d^{2}\sigma}{d\Omega_{k^{\prime}}d\epsilon^{\prime}}=\sigma_{M}\left
[\left (\frac{Q^{2}}{q^{2}}\right )^{2}R_{L}(q,\omega)+\left
(\frac{1}{2}\left|\frac{Q^{2}}{q^{2}}\right
|+\tan^{2}\frac{\theta}{2}\right )R_{T}(q,\omega)\right],
\label{eq:62}
\end{equation}
where
\begin{equation}
\sigma_{M}=\left [\frac{\alpha \cos (\theta/2)}{2\epsilon
\sin^{2}(\theta/2)}\right ]^{2}
\label{eq:63}
\end{equation}
is the Mott cross section and $\alpha$ is the fine structure
constant. In Eq.~(\ref{eq:62}) $R_{L}$ and $R_{T}$ are the
longitudinal and transverse response functions which contain all
the information on the distribution of the nuclear electromagnetic
charge and current densities, being projections (with respect to
the momentum transfer direction) of the nuclear currents. They can
be separated experimentally by plotting the cross section against
$\tan^{2}(\theta/2)$ at fixed $(q,\omega)$ (the so-called
"Rosenbluth plot"). These functions can be evaluated as components
of the nuclear tensor $W_{\mu\nu}$. In \cite{AMD+88} this tensor
is computed in the framework of the RFG model and $R_{L(T)}$ for
the QE electron scattering are expressed by means of the RFG
scaling function (Eq.~(\ref{eq:9}) of Ref.~\cite{AMD+88}).

At leading-order in the parameter $k_{F}/m_{N}$ the QE responses
have the form \cite{Amaro2005}:
\begin{equation}
R_{L}^{QE}(\kappa,\lambda)=\Lambda_{0}\frac{\kappa^{2}}{\tau}[(1+\tau)W_{2}(\tau)-
W_{1}(\tau)]\times f_{RFG}^{QE}(\psi^{\prime}),
\label{eq:64}
\end{equation}
\begin{equation}
R_{T}^{QE}(\kappa,\lambda)=\Lambda_{0}[2W_{1}(\tau)]\times
f_{RFG}^{QE}(\psi^{\prime}),
\label{eq:65}
\end{equation}
with
\begin{equation}
\Lambda_{0}\equiv \frac{{\cal N}\xi_{F}}{m_{N}\kappa\eta_{F}^{3}},
\label{eq:66}
\end{equation}
where ${\cal N}=Z$ or $N$ and $W_{1}$, $W_{2}$ are the structure
functions for elastic scattering which are linked to the Sachs
form factors
\begin{equation}
(1+\tau)W_{2}(\tau)-W_{1}(\tau)=G_{E}^{2}(\tau),
\label{eq:67}
\end{equation}
\begin{equation}
2W_{1}(\tau)=2\tau G_{M}^{2}(\tau).
\label{eq:68}
\end{equation}

In \cite{Amaro99,Amaro2005} the electro-production of the
$\Delta$-resonance is considered computing the nuclear tensor also
within the RFG model and analytical expressions for the response
functions are obtained. The latter contain the RFG $\Delta$-peak
scaling function (\ref{eq:54}) and read \cite{Amaro99}:
\begin{equation}
R_{L}(\kappa,\lambda)=\frac{3{\cal
N}\xi_{F}}{2m_{N}\eta_{F}^{3}\kappa}\frac{\kappa^{2}}{\tau}[(1+\tau\rho^{2})w_{2}(\tau)-
w_{1}(\tau)+w_{2}(\tau)D(\kappa,\lambda)]\times
f_{RFG}^{\Delta}(\psi_{\Delta}^{\prime}),
\label{eq:69}
\end{equation}
\begin{equation}
R_{T}(\kappa,\lambda)=\frac{3{\cal
N}\xi_{F}}{2m_{N}\eta_{F}^{3}\kappa}[2w_{1}(\tau)+w_{2}(\tau)D(\kappa,\lambda)]\times
f_{RFG}^{\Delta}(\psi_{\Delta}^{\prime}),
\label{eq:70}
\end{equation}
where ${\cal N}=Z$ or $N$,
\begin{eqnarray}
D(\kappa,\lambda)& \equiv &
\frac{\tau}{\kappa^{2}}[(\lambda\rho+1)^{2}+(\lambda\rho+1)(1+\psi_{\Delta}^
{\prime 2})\xi_{F}\\ \nonumber &+&
\frac{1}{3}(1+\psi_{\Delta}^{\prime 2}+\psi_{\Delta}^{\prime
4})\xi_{F}^{2}]-(1+\tau\rho^{2}).
\label{eq:71}
\end{eqnarray}
The single-baryon structure functions can be expressed by means of
the electric $(G_{E})$, magnetic $(G_{M})$ and Coulomb $(G_{C})$
delta form factors \cite{Amaro99}:
\begin{equation}
w_{1}(\tau)=\frac{1}{2}(\mu_{\Delta}+1)^{2}(2\tau\rho+1-\mu_{\Delta})
(G_{M}^{2}+3G_{E}^{2}),
\label{eq:72}
\end{equation}
\begin{equation}
w_{2}(\tau)=\frac{1}{2}(\mu_{\Delta}+1)^{2}\frac{2\tau\rho+1-\mu_{\Delta}}{1+\tau\rho^{2}}
\left
(G_{M}^{2}+3G_{E}^{2}+4\frac{\tau}{\mu_{\Delta}^{2}}G_{C}^{2}
\right ).
\label{eq:73}
\end{equation}
These form factors are parametrized as follows \cite{Amaro99}:
\begin{equation}
G_{M}(Q^{2})=2.97f(Q^{2}),
\label{eq:74}
\end{equation}
\begin{equation}
G_{E}(Q^{2})=-0.03f(Q^{2}),
\label{eq:75}
\end{equation}
\begin{equation}
G_{C}(Q^{2})=-0.15G_{M}(Q^{2}),
\label{eq:76}
\end{equation}
where
\begin{equation}
f(Q^{2})=G_{E}^{P}(Q^{2})\frac{1}{\left
[1-\frac{Q^{2}}{3.5(GeV/c)^{2}}\right]^{1/2}}
\label{eq:77}
\end{equation}
with
\begin{equation}
G_{E}^{P}=\frac{1}{(1+4.97\tau)^{2}}
\label{eq:78}
\end{equation}
being the Galster parametrization \cite{Galster71} of the electric
form factor.

In the CDFM the longitudinal and transverse response functions can
be obtained by averaging the RFG response functions in the QE
region [Eqs.~(\ref{eq:64}) and (\ref{eq:65})] and $\Delta$-region
[Eqs.~(\ref{eq:69}) and (\ref{eq:70})] by means of the weight
functions in $r$-space $|F(R)|^{2}$ and $k$-space
$|G(\overline{k}_{F})|^{2}$, similarly as in the case of the QE-
and $\Delta$-scaling functions (Eqs.~(\ref{eq:1}), (\ref{eq:5}),
(\ref{eq:15}), (\ref{eq:23}) and (\ref{eq:55}), (\ref{eq:58}),
respectively). As a result, accounting for the different behavior
of the RFG scaling functions and terms containing
$\eta_{F}(R)=k_{F}(R)/m_{N}$ as functions of $R$ or
$\overline{k}_{F}=\alpha/R$ in (\ref{eq:64}), (\ref{eq:65}),
(\ref{eq:69}) and (\ref{eq:70}), the CDFM response functions
$R_{L(T)}$ in QE- and $\Delta$-regions have approximately the same
forms as in the equations just mentioned, in which, however, the
RFG scaling functions are changed by the CDFM scaling functions
obtained in Sections 2 and 3.

\begin{figure}[htb]
\includegraphics[width=10cm]{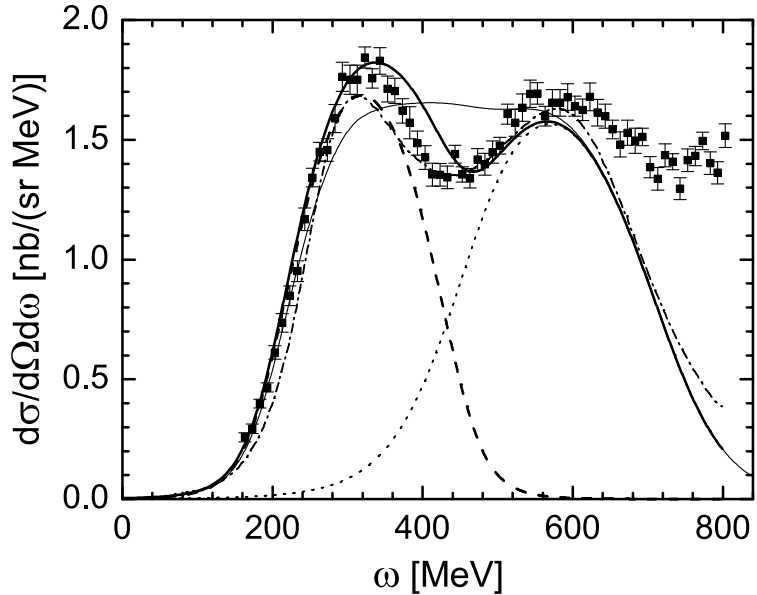}
\caption{Inclusive electron scattering on $^{12}$C at
$\epsilon=1299$ MeV and $\theta=37.5^{\circ}$ ($q_{exp}^{QE}=792$
MeV/c $>2k_{F}$). The results obtained using $c_{1}^{QE}=0.72$ in
the CDFM scaling function for the QE cross section and the total
result are given by dashed and thick solid line, respectively.
Dotted line: using CDFM $\Delta$-scaling function; thin solid
line: total CDFM result with $c_{1}^{QE}=0.63$. Dash-dotted line:
result of ERFG method \protect\cite{BCD+04,Amaro2005}. The
experimental data are taken from \protect\cite{Sealock89}.
\label{fig06}}
\end{figure}

\begin{figure}[htb]
\includegraphics[width=10cm]{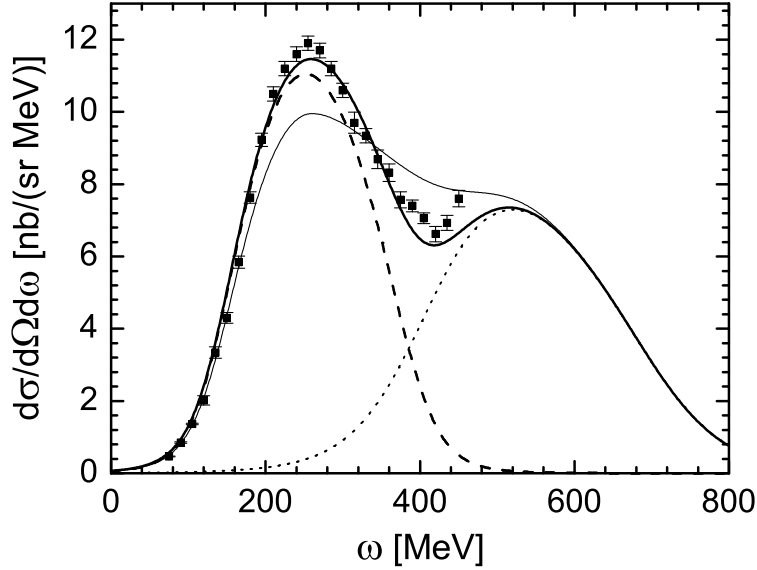}
\caption{Inclusive electron scattering on $^{12}$C at
$\epsilon=2020$ MeV and $\theta=20.02^{\circ}$ ($q_{exp}^{QE}=703$
MeV/c $>2k_{F}$). The results obtained using $c_{1}^{QE}=0.73$ in
the CDFM scaling function for the QE cross section and the total
result are given by dashed and thick solid line, respectively.
Dotted line: using CDFM $\Delta$-scaling function. Thin solid
line: total CDFM result with $c_{1}^{QE}=0.63$. The experimental
data are taken from \protect\cite{Day93}.
\label{fig07}}
\end{figure}

\begin{figure}[htb]
\includegraphics[width=10cm]{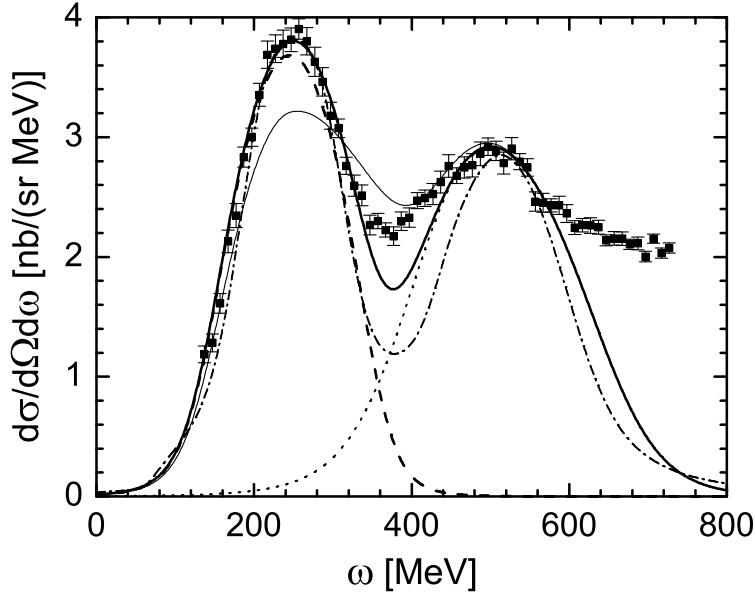}
\caption{The same as in Fig.~\ref{fig07} for $\epsilon=1108$ MeV
and $\theta=37.5^{\circ}$ ($q_{exp}^{QE}=675$ MeV/c $>2k_{F}$).
Dot-dashed line: using QE- and $\Delta$-scaling functions obtained
in the LFD approach. The experimental data are taken from
\protect\cite{Sealock89}.
\label{fig08}}
\end{figure}

\begin{figure}[htb]
\includegraphics[width=10cm]{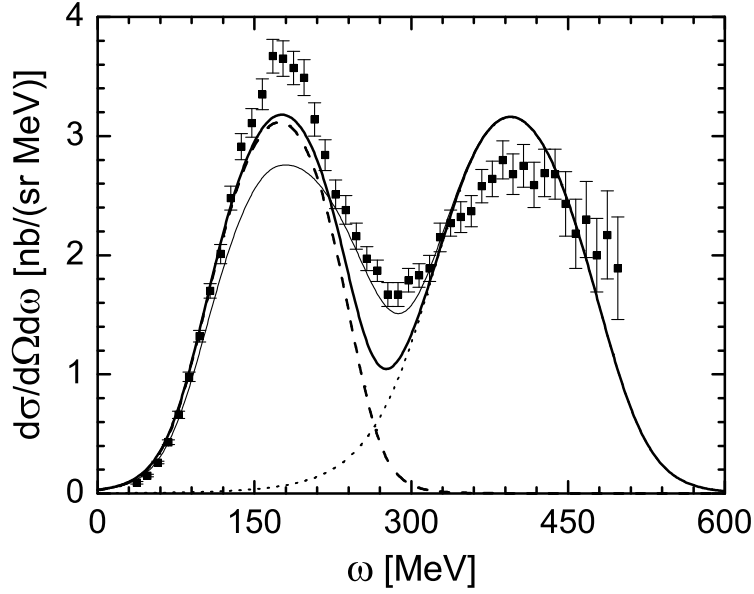}
\caption{The same as in Fig.~\ref{fig07} for $\epsilon=620$ MeV
and $\theta=60^{\circ}$ ($q_{exp}^{QE}=552$ MeV/c $>2k_{F}$). The
experimental data are taken from \protect\cite{Barr83}.
\label{fig09}}
\end{figure}

\begin{figure}[htb]
\includegraphics[width=10cm]{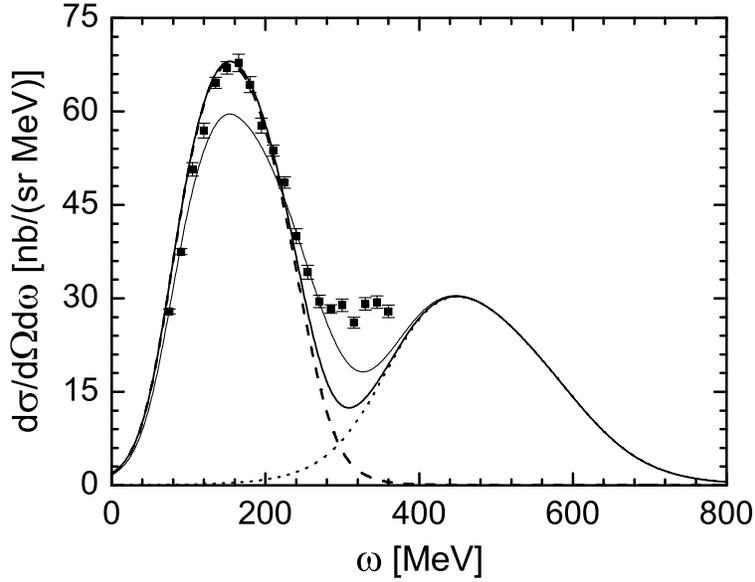}
\caption{The same as in Fig.~\ref{fig06} for $\epsilon=2020$ MeV
and $\theta=15.02^{\circ}$ ($q_{exp}^{QE}=530$ MeV/c $>2k_{F}$)
for the CDFM results. The experimental data are taken from
\protect\cite{Day93}.
\label{fig10}}
\end{figure}

\begin{figure}[htb]
\includegraphics[width=10cm]{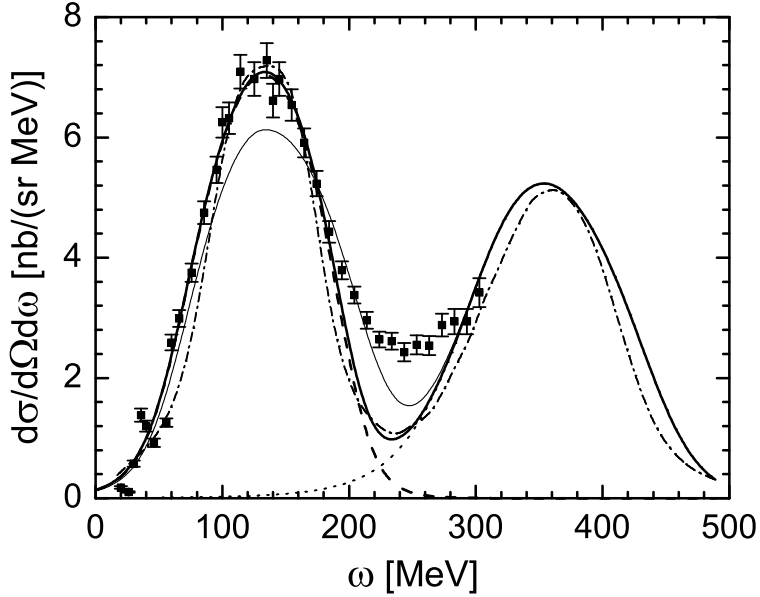}
\caption{The same as in Fig.~\ref{fig06} for $\epsilon=500$ MeV
and $\theta=60^{\circ}$ ($q_{exp}^{QE}=450$ MeV/c $\geq 2k_{F}$).
Here the dot-dashed line shows the result using QE- and
$\Delta$-scaling functions obtained in the LFD approach. The
experimental data are taken from \protect\cite{Whitney74}.
\label{fig11}}
\end{figure}

\begin{figure}[htb]
\includegraphics[width=10cm]{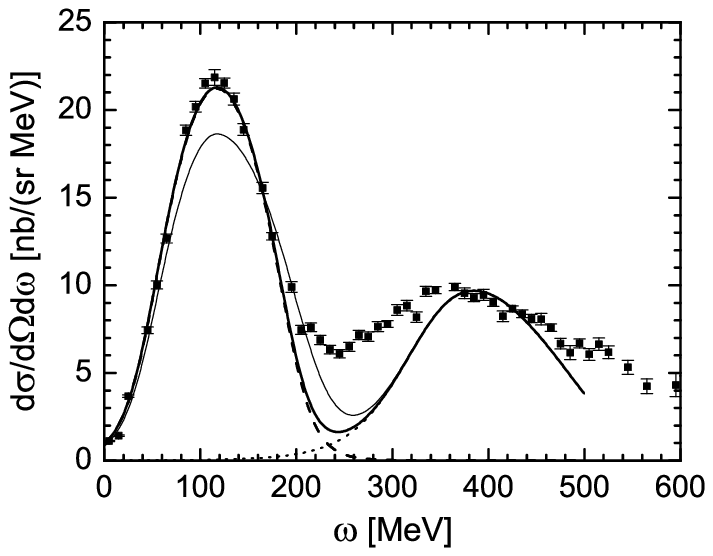}
\caption{The same as in Fig.~\ref{fig06} for $\epsilon=730$ MeV
and $\theta=37.1^{\circ}$ ($q_{exp}^{QE}=442$ MeV/c $\leq 2k_{F}$)
for the CDFM results. The experimental data are taken from
\protect\cite{Oconnell87}.
\label{fig12}}
\end{figure}

\begin{figure}[htb]
\includegraphics[width=10cm]{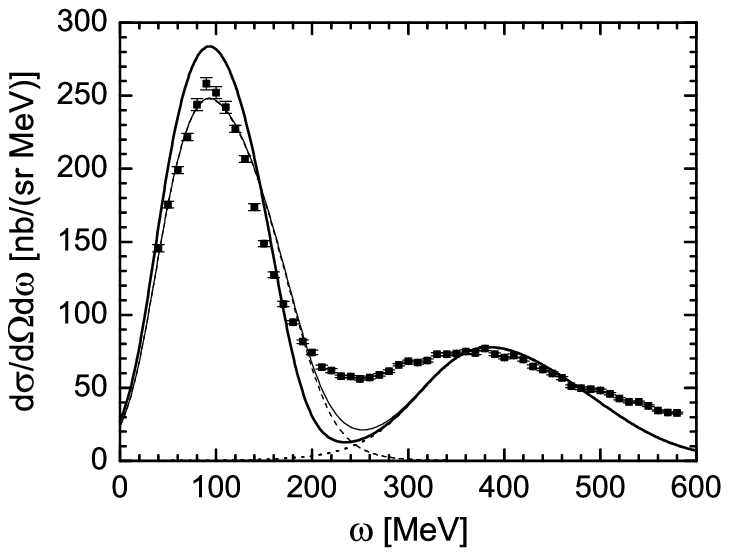}
\caption{The same as in Fig.~\ref{fig06} for $\epsilon=1650$ MeV
and $\theta=13.5^{\circ}$ ($q_{exp}^{QE}=390$ MeV/c $\leq 2k_{F}$)
for the CDFM results. The experimental data are taken from
\protect\cite{Baran88}.
\label{fig13}}
\end{figure}

\begin{figure}[htb]
\includegraphics[width=10cm]{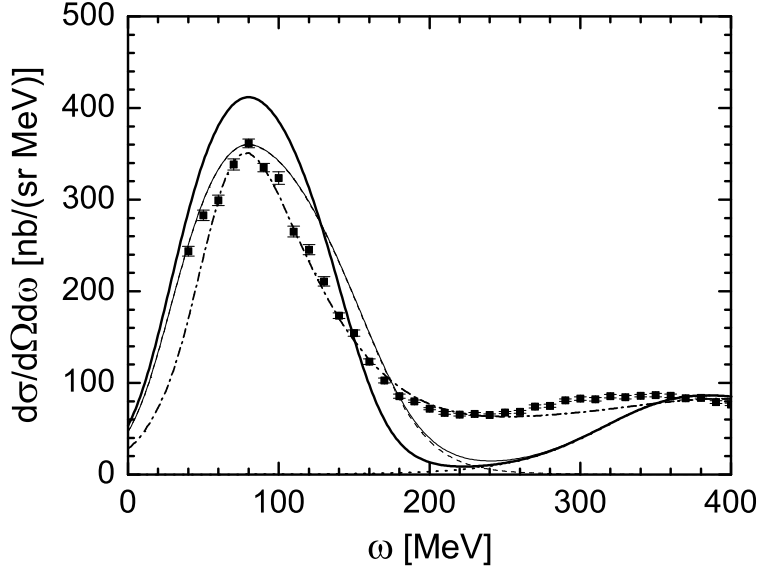}
\caption{The same as in Fig.~\ref{fig06} for $\epsilon=1500$ MeV
and $\theta=13.5^{\circ}$ ($q_{exp}^{QE}=352$ MeV/c $\leq
2k_{F}$). The experimental data are taken from
\protect\cite{Baran88}.
\label{fig14}}
\end{figure}

\begin{figure}[htb]
\includegraphics[width=10cm]{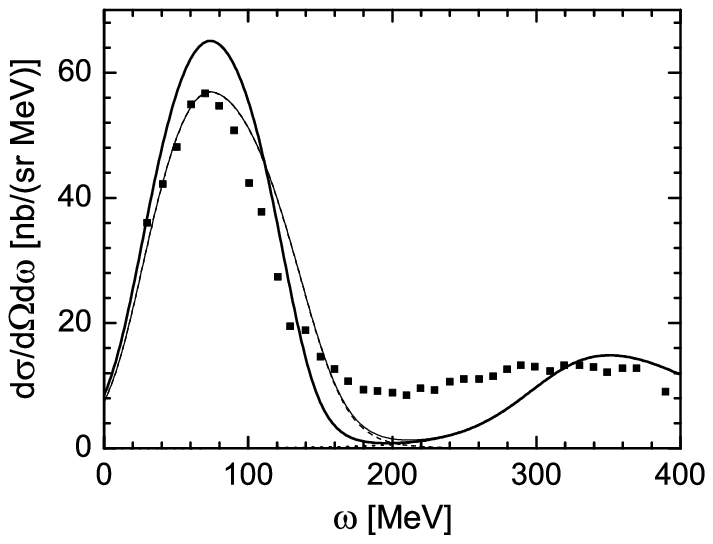}
\caption{The same as in Fig.~\ref{fig06} for $\epsilon=537$ MeV
and $\theta=37.1^{\circ}$ ($q_{exp}^{QE}=326$ MeV/c $\leq 2k_{F}$)
for the CDFM results. The experimental data are taken from
\protect\cite{Oconnell87}.
\label{fig15}}
\end{figure}

In Figs.~\ref{fig06}-\ref{fig15} we give results of calculations
within the CDFM of inclusive electron scattering on $^{12}$C at
different incident energies and angles. The QE-contribution is
calculated using the Fermi-type density distribution of $^{12}$C
with the same values of the parameters as in \cite{AGK+04,AGK+05}:
$R=2.47$ fm and $b=0.42$ fm (which lead to a charge rms radius
equal to the experimental one) and Fermi momentum $k_{F}=1.156$
fm$^{-1}$. The delta-contribution is calculated using the
necessary changes of the parameter values of the Fermi-type
density (used in Fig.~\ref{fig05}): $R_{\Delta}=1.565$ fm,
$b_{\Delta}=0.42$ fm and $k_{F}=1.20$ fm$^{-1}$. The coefficient
$c_{1}$ used in the $\Delta$-region scaling function is fixed to
be equal to 0.54 so that the maximum of the scaling function to be
in agreement with the data. The scaling function
$f^{\Delta}(\psi_{\Delta}^{\prime})$ is symmetric, its maximum is
chosen to be 0.54 (but not 0.75) and it is normalized to unity by
means of the fixed value of $k_{F}=1.20$ fm$^{-1}$. The inclusive
electron-$^{12}$C scattering cross sections shown in
Figs.~\ref{fig06}-\ref{fig15} are the sum of the QE and
$\Delta$-contribution. The results of the CDFM calculations are
presented for two values of the coefficient $c_{1}$ in the QE case
(noted further by $c_{1}^{QE}$), namely for $c_{1}^{QE}\simeq
0.72$ and $c_{1}^{QE}=0.63$. This is related to two types of
experimental data. In the first one the transferred momentum in
the position of the maximum of the QE peak extracted from data
($\omega_{exp}^{QE}$) is $q_{exp}^{QE}\geq 450$ MeV/c $\approx
2k_{F}$, roughly corresponding to the domain where scaling is
fulfilled \cite{BCD+04,Amaro2005}. Such cases are presented in
Figs.~\ref{fig06}-\ref{fig12}. In these cases we found by fitting
to the maximum of the QE peak the value of $c_{1}^{QE}$ to be
0.72--0.73, i.e. it is not the same as in the RFG model case (case
of symmetry of the RFG and of the CDFM scaling functions with
$c_{1}^{QE}=0.75$), but is slightly lower. This leads to a weak
asymmetry of the CDFM scaling function for cases in which
$q_{exp}^{QE}\geq 450$ MeV/c. In the second type of experimental
data $q_{exp}^{QE}$ is not in the scaling region
($q_{exp}^{QE}<450$ MeV/c). Such cases are given in
Figs.~\ref{fig13}-\ref{fig15}. For them we found by fitting to the
maximum of the QE peak extracted from data the value of
$c_{1}^{QE}$ to be 0.63. For these cases the CDFM scaling function
is definitely asymmetric. So, the results in
Figs.~\ref{fig06}-\ref{fig15} are presented for both almost
symmetric ($c_{1}^{QE}\simeq 0.72$) and asymmetric
($c_{1}^{QE}=0.63$) CDFM scaling functions. One can see that the
results for the almost symmetric CDFM scaling function agree with
the electron data in the region close to the QE peak in cases
where $q_{exp}^{QE}\geq 450$ MeV/c and overestimate the data for
cases where approximately $q_{exp}^{QE}<450$ MeV/c. The results
with asymmetric CDFM scaling function agree with the data in cases
where $q_{exp}^{QE}<450$ MeV/c and underestimate the data in cases
where $q_{exp}^{QE}\geq 450$ MeV/c. Here we would like to
emphasize that, in our opinion, the usage of asymmetric CDFM
scaling function is preferable, though the results in some cases
can underestimate the empirical data, because other additional
effects, apart from QE and $\Delta$-resonance (e.g. meson exchange
currents effects) could give important contributions to the cross
section for some specific kinematics and minor for others. A
similar situation occurs for the results obtained within the RMF
approach \cite{Caballero2006} particularly when the CC2 current
operator is selected.

In Table~\ref{table1} we list the energies, the angles, the values
of $c_{1}^{QE}$ obtained by fitting the magnitude of the QE peak,
and the energy shifts in the QE and $\Delta$-case, as well as the
approximate values of the transfer momentum $q_{exp}^{QE}$ in the
position of the maximum of the QE peak ($\omega_{exp}^{QE}$) for
different cases. The values of the energy shifts
$\epsilon_{shift}^{QE(\Delta)}$ for the QE- and $\Delta$-regions
are generally between 20 and 30 MeV. In the Figures we also
present the QE-contribution (as well as $\Delta$-contribution) for
the value of $c_{1}^{QE}$ which fits approximately the magnitude
of the QE peak.

\begin{table}[h]
\caption{Values of energies $\epsilon$, angles $\theta$, the
coefficient $c_{1}^{QE}$ obtained by fitting the magnitude of the
QE peak, energy shifts $\epsilon_{shift}^{QE}$ and
$\epsilon_{shift}^{\Delta}$, and transferred momenta
$q_{exp}^{QE}$ for the cases of inclusive electron scattering
cross sections considered. Energies are in MeV, angles are in
degrees and momenta are in MeV/c.}
{\begin{tabular}{cccccccccccccc} \hline \hline Figure & &
$\epsilon$ & & $\theta$ & & $c_{1}^{QE}$ & &
$\epsilon_{shift}^{QE}$ & & $\epsilon_{shift}^{\Delta}$ & & $\approx q_{exp}^{QE}$ \\
\hline
6     & & 1299  & & 37.5  & & 0.72 & & 30 & & 30 & & 792  \\
7     & & 2020  & & 20.02 & & 0.73 & & 25 & & 20 & & 703  \\
8     & & 1108  & & 37.5  & & 0.73 & & 30 & & 30 & & 675  \\
9     & & 620   & & 60    & & 0.73 & & 20 & & 0  & & 552  \\
10    & & 2020  & & 15.02 & & 0.72 & & 20 & & 30 & & 530  \\
11    & & 500   & & 60    & & 0.72 & & 30 & & 0  & & 450  \\
12    & & 730   & & 37.1  & & 0.72 & & 20 & & 20 & & 442 $\simeq 2k_{F}$  \\
13    & & 1650  & & 13.5  & & 0.63 & & 20 & & 30 & & 390  \\
14    & & 1500  & & 13.5  & & 0.63 & & 20 & & 20 & & 352  \\
15    & & 537   & & 37.1  & & 0.63 & & 20 & & 20 & & 326  \\
\hline \hline
\end{tabular}}
\label{table1}
\end{table}

In Figs.~\ref{fig08} and \ref{fig11} we present also the
calculations of the electron cross sections using QE- and
$\Delta$-scaling functions obtained by using the nucleon momentum
distributions obtained in the LFD approach (Section 3) which give
a reasonable agreement with the empirical electron scattering
data. In Figs.~\ref{fig06} and \ref{fig14} we also give for
comparison the results of the cross sections obtained within the
ERFG method \cite{BCD+04,Amaro2005}. In this method the response
functions and differential cross sections are calculated using the
scaling function fitted to the data.

It is interesting to note that for those kinematics where the
overlap between the QE and $\Delta$ peaks is bigger
(Figs.~\ref{fig06}, \ref{fig07} and \ref{fig08}), the asymmetric
CDFM model ($c_{1}^{QE}=0.63$) gives rise to an excess of strength
in the transition region. This makes a difference with the ERFG
model (see Fig.~\ref{fig06}) which fits nicely the data in that
region. This discrepancy between the two models, asymmetric CDFM
and ERFG, can be explained by noting the different behavior
presented by the two scaling functions in the region of
$\psi^{\prime}$ between 0.5 and 1.5, being the asymmetric CDFM one
significantly larger.

Note on the other hand, that in the cases where the overlap
between the QE and $\Delta$ peaks is weaker
(Figs.~\ref{fig09}-\ref{fig12}), the asymmetric CDFM model,
compared to the almost symmetric CDFM one, reproduces better the
data in the transition region although it underpredicts
importantly the maximum of the QE peak. Concerning results in
Figs.~\ref{fig13}-\ref{fig15} (it can be also applied to
Figs.~\ref{fig11} and \ref{fig12}), one observes that both CDFM
approaches do not reproduce the strength of data located in the
region between the QE and delta peaks. This is not the case for
the ERFG model (see Fig.~\ref{fig14}) which fits nicely the
experiment for $\omega\geq 180$ MeV. This result is connected with
the much bigger strength of the scaling function provided by the
ERFG model for larger values of the scaling variable,
$\psi^{\prime}\geq 2$ (see Fig.~\ref{fig02}).

From this whole analysis, one may conclude that the
phenomenological procedure introduced in the CDFM model to get an
asymmetric scaling function, gives rise to an excess of strength
in the region $0.5\leq\psi^{\prime}\leq 1.5$, whereas the model
lacks strength for larger $\psi^{\prime}$-values,
$\psi^{\prime}\geq 2$.

\subsection{Scaling functions and charge-changing neutrino-nucleus
reaction cross sections}

In this Subsection we will present applications of the CDFM and
LFD scaling functions to calculations of charge-changing
neutrino-nucleus reaction cross sections. We follow the
description of the formalism given in \cite{Amaro2005}. The
charge-changing neutrino cross section in the target laboratory
frame is given in the form
\begin{equation}
\left [ \frac{d^{2}\sigma}{d\Omega dk^{\prime}}\right
]_{\chi}\equiv \sigma_{0}{\cal F}_{\chi}^{2},
\label{eq:79}
\end{equation}
where $\chi=+$ for neutrino-induced reaction (e.g.,
$\nu_{\ell}+n\rightarrow \ell^{-}+p$, where $\ell=e, \mu, \tau$)
and $\chi=-$ for antineutrino-induced reactions (e.g.,
$\overline{\nu}_{\ell}+p\rightarrow \ell^{+}+n$),
\begin{equation}
\sigma_{0}\equiv
\frac{(G\cos\theta_{c})^{2}}{2\pi^{2}}[k^{\prime}\cos\tilde{\theta}/2]^{2},
\label{eq:80}
\end{equation}
$G=1.16639\times 10^{-5}$ GeV$^{-2}$ being the Fermi constant,
$\theta_{c}$ being Cabibbo angle $(\cos\theta_{c}=0.9741)$,
\begin{equation}
\tan^{2}\tilde{\theta}/2\equiv \frac{|Q|^{2}}{v_{0}},
\label{eq:81}
\end{equation}
\begin{equation}
v_{0}\equiv (\epsilon +
\epsilon^{\prime})^{2}-q^{2}=4\epsilon\epsilon^{\prime}-|Q|^{2}.
\label{eq:82}
\end{equation}

The quantity ${\cal F}_{\chi}^{2}$ which depends on the nuclear
structure is written in \cite{Amaro2005} as a generalized
Rosenbluth decomposition having charge-charge,
charge-longitudinal, longitudinal-longitudinal and two types of
transverse responses. The nuclear response functions are expressed
in terms of the nuclear tensor $W^{\mu\nu}$ in both QE and
$\Delta$-regions using its relationships with the RFG model
scaling functions. Following \cite{Amaro2005}, in the calculations
of the neutrino-nucleus cross sections the H\"{o}hler
parametrization 8.2 \cite{Hohler76} of the form factors in the
vector sector was used, while in the axial-vector sector the form
factors given in \cite{Amaro2005} were used.

\begin{figure}[htb]
\includegraphics[width=15cm]{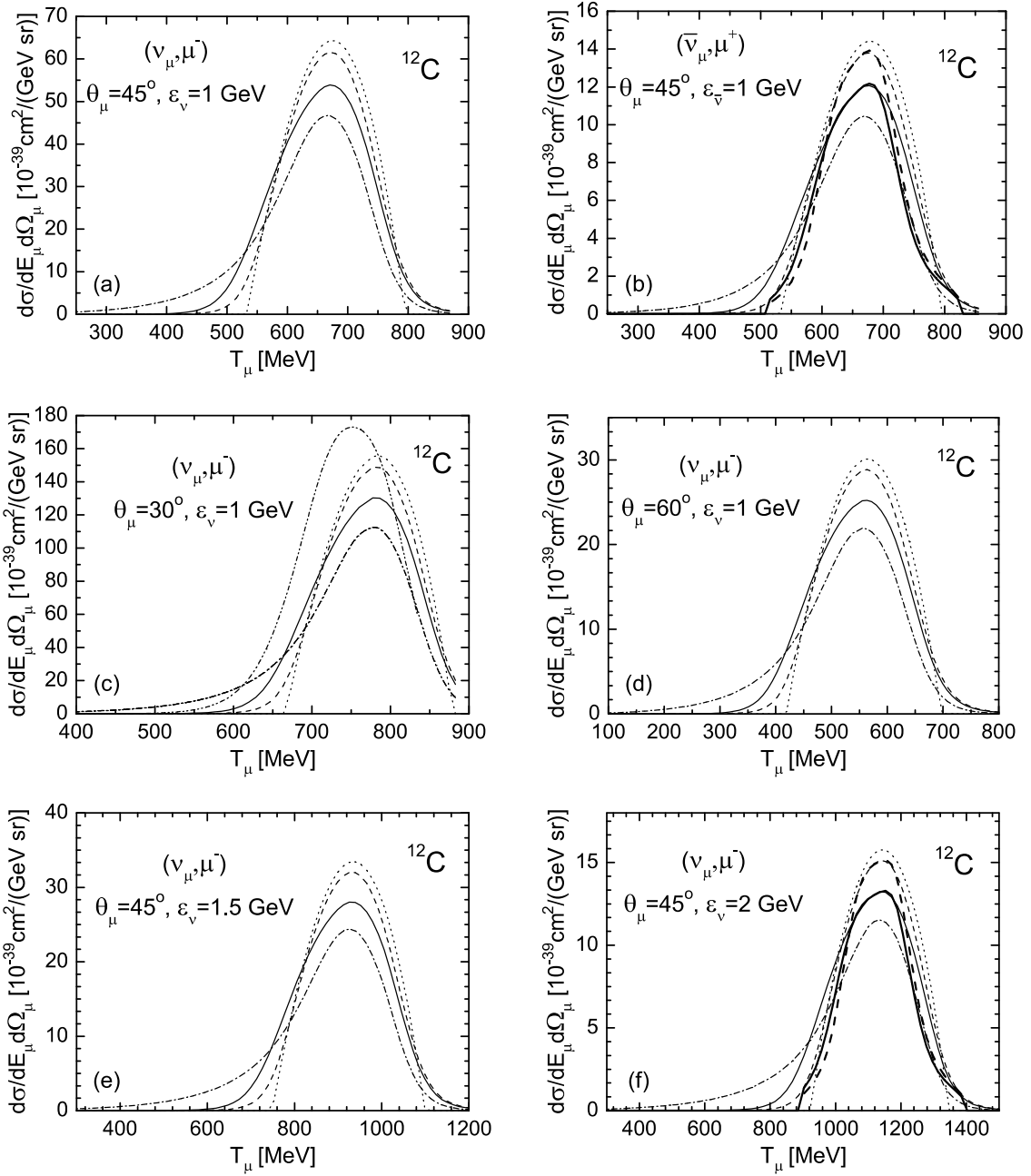}
\caption{The cross section of quasielastic charge-changing
$(\nu_{\mu},\mu^{-})$ reaction [(a), (c)-(f)] and of
$(\overline{\nu}_{\mu},\mu^{+})$ reaction (b) on $^{12}$C for
$\epsilon=1, 1.5$ and 2 GeV using QE-scaling functions in CDFM
(thin solid line: with $c_{1}=0.63$; thin dashed line: with
$c_{1}=0.72$). The results using QE-scaling functions in LFD
(thick solid line: with $c_{1}=0.63$; thick dashed line: with
$c_{1}=0.72$) are presented in (b) and (f). The RFG model result
and ERFG result \protect\cite{BCD+04,Amaro2005} are shown by
dotted and dash-dotted lines, respectively.
\label{fig16}}
\end{figure}

In our work, instead of the RFG scaling functions in QE- and
$\Delta$-regions, we use those obtained in the CDFM and LFD
approach (Sections 2 and 3). In Fig.~\ref{fig16} we give the
results of calculations for cross sections of QE neutrino
$(\nu_{\mu},\mu^{-})$ scattering (Figs.~\ref{fig16}a, c, d, e, f)
on $^{12}$C and also antineutrino $(\overline{\nu}_{\mu},\mu^{+})$
scattering (Fig.~\ref{fig16}b) for energies of neutrino
$\epsilon_{\nu}=1, 1.5$ and 2 GeV and of antineutrino
$\epsilon_{\overline{\nu}}=1$ GeV. The presented cross sections
are functions of muon kinetic energy. The energy shift is equal to
20 MeV. The calculations of the neutrino-nucleus cross sections in
the $\Delta$-region will be a subject of a future work.

We give the results of our calculations using the CDFM scaling
function which is almost symmetric (with $c_{1}=0.72$), as well as
the asymmetric CDFM scaling function (with $c_{1}=0.63$). These
values of $c_{1}$ correspond to the cases of inclusive electron
scattering considered. As can be seen the results obtained by
using the almost symmetric CDFM scaling function are close to the
RFG model results. On the other hand, the results obtained with
the use of asymmetric CDFM and LFD scaling functions are quite
different from those in the RFG model, but are close to the
predictions of the ERFG model \cite{BCD+04,Amaro2005}. The basic
difference from the ERFG model result is observed in the tail
extended to small muon energy values, where the ERFG model gives
more strength.

\section{Conclusions}

The results of the present work can be summarized as follows:

i) In Ref.~\cite{AGK+05} we extended the CDFM description of the
quasielastic $\psi^{\prime}$-scaling function from \cite{AGK+04}
by expressing it explicitly and equivalently by means of both
density and nucleon momentum distributions. In
\cite{AGK+04,AGK+05} our results on $f^{QE}(\psi^{\prime})$ were
obtained on the basis of the experimental data on the charge
densities for a wide range of nuclei. In the present work we
extended our approach to consider the scaling function
$f^{QE}(\psi^{\prime})$ for medium and heavy nuclei with $Z\neq N$
for which the proton and neutron densities are not similar. In
this case $f^{QE}(\psi^{\prime})$ is a sum of the proton and
neutron scaling functions calculated by means of the proton and
neutron densities obtained from nonrelativistic self-consistent
mean-field calculations. This concerns calculations, as examples,
of nuclei like $^{197}$Au, $^{82}$Kr, as well as $^{62}$Ni and
$^{118}$Sn \cite{AIG+06}. The comparison with the data from
\cite{DS99l,DS99} shows superscaling for negative values of the QE
$\psi^{\prime}$ including $\psi^{\prime}<-1$, whereas in the RFG
model $f(\psi^{\prime})=0$ for $\psi^{\prime}\leq -1$ (see
Fig.~\ref{fig01}).

ii) We introduce the asymmetry in the CDFM QE scaling function
using the fact that the maximum value of $f(\psi^{\prime})$ in RFG
model is 3/4 while the empirical scaling function reaches values
smaller than 0.6. In relation with this and the normalization, we
parametrize the RFG scaling function for $\psi^{\prime}\geq 0$,
thus simulating the role of all the effects which lead to
asymmetry and imposing this to the CDFM  QE scaling function. In
this way, simulating phenomenologically the effects which violate
the symmetry of $f^{QE}(\psi^{\prime})$ for $\psi^{\prime}\geq 0$
including the role of the FSI, one can obtain in the CDFM a
reasonable agreement of $f^{QE}(\psi^{\prime})$ with the empirical
data also for positive values of $\psi^{\prime}$
(Fig.~\ref{fig02}).

iii) We obtain the QE scaling function also on the basis of
calculations of nucleon momentum distribution $n(k)$ within an
approach based on the light-front dynamics method
\cite{AGI+02,CK95,CDK98} which improves that used in
Ref.~\cite{AGK+05}. Here we include in the particle-state part of
$n(k)$ not only a contribution of the function $f_{5}$ (as in
\cite{AGI+02} and \cite{AGK+05}) but also a contribution of the
function $f_{2}$. $f_{5}$ and $f_{2}$ are two of the six scalar
components of the deuteron total wave function in the LFD
\cite{CK95,CDK98} and are the main contributions to the tail of
$n_{d}(k)$. It can be seen in Fig.~\ref{fig03} the reasonable
agreement of $n(k)$ in LFD with the $y_{CW}$-scaling data
\cite{CW99,CW97}. This result made it possible to obtain a good
description of the experimental QE scaling function
(Fig.~\ref{fig04}) at least up to $\psi^{\prime}\simeq -1.2$.

iv) We extend our analysis within the CDFM and LFD to the
$\Delta$-peak region which is the main contribution to the
inelastic scattering. Here we emphasize that reasonable agreement
with the experimental data (Fig.~\ref{fig05}) was obtained using
the empirical value of the coefficient in front of the RFG scaling
function (0.54 instead of 0.75) in both CDFM and LFD. Also, the
parameter $R_{\Delta}$ used in the Fermi-type density for $^{12}$C
(necessary to calculate the weight function $|F_{\Delta}(R)|^{2}$
and thus the scaling function $f^{\Delta}
(\psi_{\Delta}^{\prime})$) has a smaller value (1.565 fm) than
that ($R$=2.42 fm) in the QE case, while the value of the
diffuseness parameter $b_{\Delta}$ remains the same as $b$ in the
QE case. We note that the use of $n_{LFD}(k)$ with the same values
of $\beta$ and of $k_{F}$ ($\beta=0.80$, $k_{F}=1.20$ fm$^{-1}$)
gives a reasonable agreement with results for both QE- and
$\Delta$-region scaling functions (Figs.~\ref{fig04} and
\ref{fig05}).

v) The QE- and $\Delta$-region scaling functions obtained in the
CDFM and in the LFD approach are applied to description of
experimental data on differential cross sections of inclusive
electron scattering by $^{12}$C at large energies and transferred
momenta (Figs.~\ref{fig06}-\ref{fig15}). The CDFM results are
presented for both almost symmetric ($c_{1}^{QE}\simeq 0.72$) and
asymmetric ($c_{1}^{QE}=0.63$) scaling functions. We observe that
there are two regions of the value of $q_{exp}^{QE}$ in different
experiments at which the above mentioned (almost symmetric and
asymmetric) scaling functions work better. The almost symmetric
scaling function leads to results in agreement with the data in
the region of the QE peak in cases when the transferred momentum
$(q_{exp}^{QE})$ in the position of maximum of the QE peak
$(\omega_{exp}^{QE})$ is in the scaling region ($q_{exp}^{QE}\geq
450$ MeV/c $\approx 2k_{F}$), while the data are overestimated in
cases where $q_{exp}^{QE}<450$ MeV/c. The results obtained when
asymmetric scaling function ($c_{1}^{QE}=0.63$) is used agree with
the data in cases when $q_{exp}^{QE}< 450$ MeV/c and underestimate
them when $q_{exp}^{QE}\geq 450$ MeV/c in the region close to the
QE peak, but differences emerge in the transition region. In our
opinion, the latter case is preferable because additional effects
(apart from QE and $\Delta$-resonance), e.g. of the meson exchange
currents could give additional important contributions to the
inclusive electron cross sections for some specific kinematics and
minor for others.

vi) The CDFM and LFD scaling functions are applied to calculations
of QE charge-changing neutrino-nuclei reaction cross sections. We
present in Fig.~\ref{fig16} the predicted cross sections for the
reactions $(\nu_{\mu},\mu^{-})$ and
$(\overline{\nu}_{\mu},\mu^{+})$ on the $^{12}$C nucleus for
energies of the incident particles from 1 to 2 GeV. Our results
are compared with those from the RFG model and from the ERFG model
\cite{BCD+04,Amaro2005}. The results obtained by using the
asymmetric CDFM scaling function are close to those of ERFG and
are quite different from the RFG results while the almost
symmetric CDFM scaling function leads to cross sections which are
similar to the results of the RFG model.

\begin{acknowledgments}
Four of the authors (A.N.A., M.V.I., M.K.G. and M.B.B.) are
grateful to C. Giusti and A. Meucci for the discussion. This work
was partly supported by the Bulgarian National Science Foundation
under Contracts No.$\Phi$-1416 and $\Phi$-1501 and by funds
provided by DGI of MCyT (Spain) under Contract Nos. FIS
2005-00640, BFM 2003-04147-C02-01, INTAS-03-54-6545, FPA
2005-04460, and FIS 2005-01105.
\end{acknowledgments}

\end{document}